\documentclass[modern]{aastex63} 

\lefthead{Kawaguchi \& Wada}
\righthead{Mass accretion in galaxy merger}

\newcommand{\bea}{\begin{eqnarray} }
\newcommand{\eea}{\end{eqnarray}}

\begin{document}


\title{Mass accretion toward black holes in the final phase of galaxy mergers}

\author{Takeru Kawaguchi}
\affiliation{Kagoshima University, Graduate School of Science and Engineering, Kagoshima 890-0065, Japan}

\author{Naomichi Yutani}
\affiliation{Kagoshima University, Graduate School of Science and Engineering, Kagoshima 890-0065, Japan}

\author{%
Keiichi Wada
}
\affiliation{Kagoshima University, Graduate School of Science and Engineering, Kagoshima 890-0065, Japan}
\affiliation{Ehime University, Research Center for Space and Cosmic Evolution, Matsuyama 790-8577, Japan}
\affiliation{Hokkaido University, Faculty of Science, Sapporo 060-0810, Japan}
\correspondingauthor{Keiichi Wada}
\email{wada@astrophysics.jp}

%


%


\begin{abstract}
We studied the final phases of galactic mergers, focusing on 
interactions between supermassive black holes (SMBHs)
and the interstellar medium in a central sub-kpc region, using an N-body/hydrodynamics code.
This numerical experiment aims to understand the fate of the gas supplied by
mergers of two or more galaxies with SMBHs, whose masses are $10^7 M_\odot$.
We observed that the mass accretion rate to one SMBH
exceeds the Eddington accretion rate when the distance between two black holes (BHs) rapidly decreases.
However, this rapid accretion phase does not last for more than $10^7$ yrs, and it
drops to $\sim$ 10\% of the Eddington rate in the quasi-steady accretion phase.
The rapid accretion is caused by the angular momentum transfer from the gas to
the stellar component, and the moderate accretion
in the quasi-steady phase is caused {by turbulent viscosity and gravitational torque }
 in the disk.
The second merger event enhances the mass accretion to the BHs;
however, this phase takes place on a similar timescale to the first merger event.
{We also found that 
the AGN feedback and the mass accretion to BHs can coexist in the central region of merged galaxies, 
 if the amount of feedback energy is given as $(2 \times 10^{-4} - 2 \times 10^{-3} )\dot{M} c^2$, where 
 $\dot{M}$ is the accretion rate to $r= 1$ pc. }
The accretion rate is suppressed by $\sim$ 1/50 in the quasi-steady accretion phase for $0.02 \dot{M} c^2$. 
The fraction of the gas that finally falls to each BH is approximately 5--7\% of the supplied total gas mass ($10^8 M_\odot$),
 and 15--20\% of the gas forms a circumnuclear gas inside 100 pc. 
This remnant gas heavily obscures the luminous phase of the active galactic nuclei (AGN) during merger events, 
and the moderate AGN feedback does not alter this property.
\end{abstract}


\section{Introduction} \label{sec:intro}

It is widely believed that galaxy mergers contribute to the growth of supermassive black holes (SMBH) in galaxies, and {nuclear activities can be triggered by major mergers} \citep{treister2012, Lanzuisi2015, Koss2016, goulding2018b}.
{However, observational studies also suggest that ``major'' mergers of galaxies might not be a sufficient condition for
triggering active galactic nuclei (AGNs). For example, 
\citet{gabor2009} analyzed the host
morphologies of AGNs selected in the cosmic evolution
survey (COSMOS) and found that
the disturbed fraction among active and quiescent galaxies at $z \sim 1$ is not significantly different. \citet{kocevski2012} also claimed that the hosts of moderate-luminosity AGNs are no more likely to be involved in an ongoing merger or interaction relative to non-active galaxies at $1.5 < z < 2.5$. These results might be due to a time lag between merger events and ignition of AGNs, but 
\citet{cisternas2011} claimed that this is unlikely based on the HST imaging analysis of AGN host galaxies in the COSMOS field.
One possible reason why it is difficult to find a direct connection between galaxy mergers and AGN activities is that 
we still do not fully understand the final phase of the galaxy merging process, especially in the central part of galaxies, and how AGN activities
are triggered.
}

{Many numerical simulations have shown that galaxy mergers enhance the mass inflow to the central region of merging galaxies \citep[e.g.,][]{Barnes1991, mihos1996, Springel2003, Springel2005, dimatteo2005, Hopkins2006, angles2017}.}
However, the detailed mass accretion and feedback processes in the regions within 100-pc of the center
are not fully resolved in cosmological simulations \citep{Hopkins2011}. Therefore,
subgrid physics regarding mass accretion, such as the Bondi accretion, are necessary for galaxy formation
simulations \citep[][references therein] {Negri2018}.
In reality, merger-driven mass accretion is neither spherical nor steady, and transfer of the angular momentum
of the gas should be considered \citep{Negri2018, angles2017}.  {In the final phase of galaxy-galaxy mergers, 
SMBHs in host galaxies may play an important role in the dynamics of the central part of the merging system through 
their non-axisymmetric potential and energy/momentum feedback to the gas \citep[e.g.,][]{tremmel2018}.}

{Mergers of two or more nucleated galaxies result in the formation of binary BHs. \citet{governato1994} studied the
orbital decay of binary black holes (BBH) during galaxy mergers using $N$-body experiments. It has also been observed that the interstellar medium (ISM) around binary BHs may affect the orbital decay of BBHs \citep{armitage2002, delvalle2012}. 
The effect of gaseous dissipation on the dynamics of BBHs and merging galaxies has also been studied \citep{dotti2007, mayer2007,kazantzidis2005}.
}
Because of the interaction between BBHs and the ISM, the nuclear starburst could be enhanced \citep{taniguchi1996}.
\citet{Debuhr2011} also studied the merger of two disk galaxies with SMBHs using {smoothed particle hydrodynamic} (SPH) simulations and
found that the BH mass accretion is self-regulated by a balance between the radiation force and the gravity of the host galaxy.
Because their model disks are on the galactic scale, the gas dynamics and mass accretion process in the central 100-pc regions are still not clear.
More recently, \citet{Prieto2017} studied the mass transport 
in high-z galaxies and BH growth
using cosmological hydrodynamic simulations, where the maximum spatial resolution was 5 pc. In their study, energy feedback from the nucleus was also implemented at approximately 20 pc.
 However, because they followed the evolution of a single BH seed, whose mass is $10^4 M_\odot$,
the effects of BBHs on the gas dynamics in the nuclear region were still not evident.
\citet{Souza2017} studied the orbital decay of a pair of SMBHs in a multiphase non-homogeneous circumnuclear disk and
found that gas-clump BH interactions can lead to faster decay of the BH's orbit. However, their feedback model was 
based on the Bondi accretion, which would not give a correct mass accretion rate under a nonaxisymmetric interaction 
between a BBH and the gas/stars in the central region.
\citet{taniguchi1996} did not include the effect of the AGN; 
{therefore, the effect of the energy feedback from AGNs on the gas dynamics
in the central region of merging galaxies is still not clear.}

In this paper, we focus on
the final merger process of the central core of merging galaxies.
Instead of performing cosmological simulations of galaxy formation, we
conduct a series of numerical experiments of mergers of BH systems 
containing gas and stars. {We use the $N$-body/SPH code
ASURA \citep{saitoh2008, saitoh2009, saitoh2013}. }
We investigate the mass accretion rates and dynamic evolution of BHs.
We also explore how the accretion process differs depending on the efficiency of the energy feedback from AGNs on a parsec scale, 
which is an essentially free parameter in more realistic numerical simulations of galaxy formation. 

{AGNs are often buried in gas, especially in ultraluminous infrared galaxies \citep[e.g.,][]{ramos-almeida2017}, and the fraction of
Compton-thick AGNs in late merger galaxies
is higher than in local hard X-ray selected AGNs \citep{ricci2017}. \citet{buchner2015} also showed that 
Compton-thick AGNs account for approximately 40\% of $\sim$ the 2000 AGNs surveyed.
It is known that the obscured fraction of AGNs depends on the AGN luminosity \citep[][reference therein]{wada2015}.
However, the observed column density itself does not tell us where the heavy obscuration is caused.
We also do not know the origin of the obscured gas, e.g., whether it is produced by past mergers.
In this paper, we also discuss these points by analyzing our numerical results.}

The structure of this paper is as follows. In \S 2.1, we describe the numerical model setup and initial conditions.
In addition to single-merger models, where one merger event is calculated, we also explore a secondary merger
event. AGN feedback is also implemented (\S 2.2).
{The numerical results for the single-merger and multiple-merger cases are shown in \S 3.}
In \S 4, we discuss nuclear obscuration and AGN activities during the merger process. We summarize the results in \S 5.

\section{Models and Methods} \label{sec:style}
\subsection{Models}

\begin{figure}[h]
\centering
\includegraphics[width = 10cm]{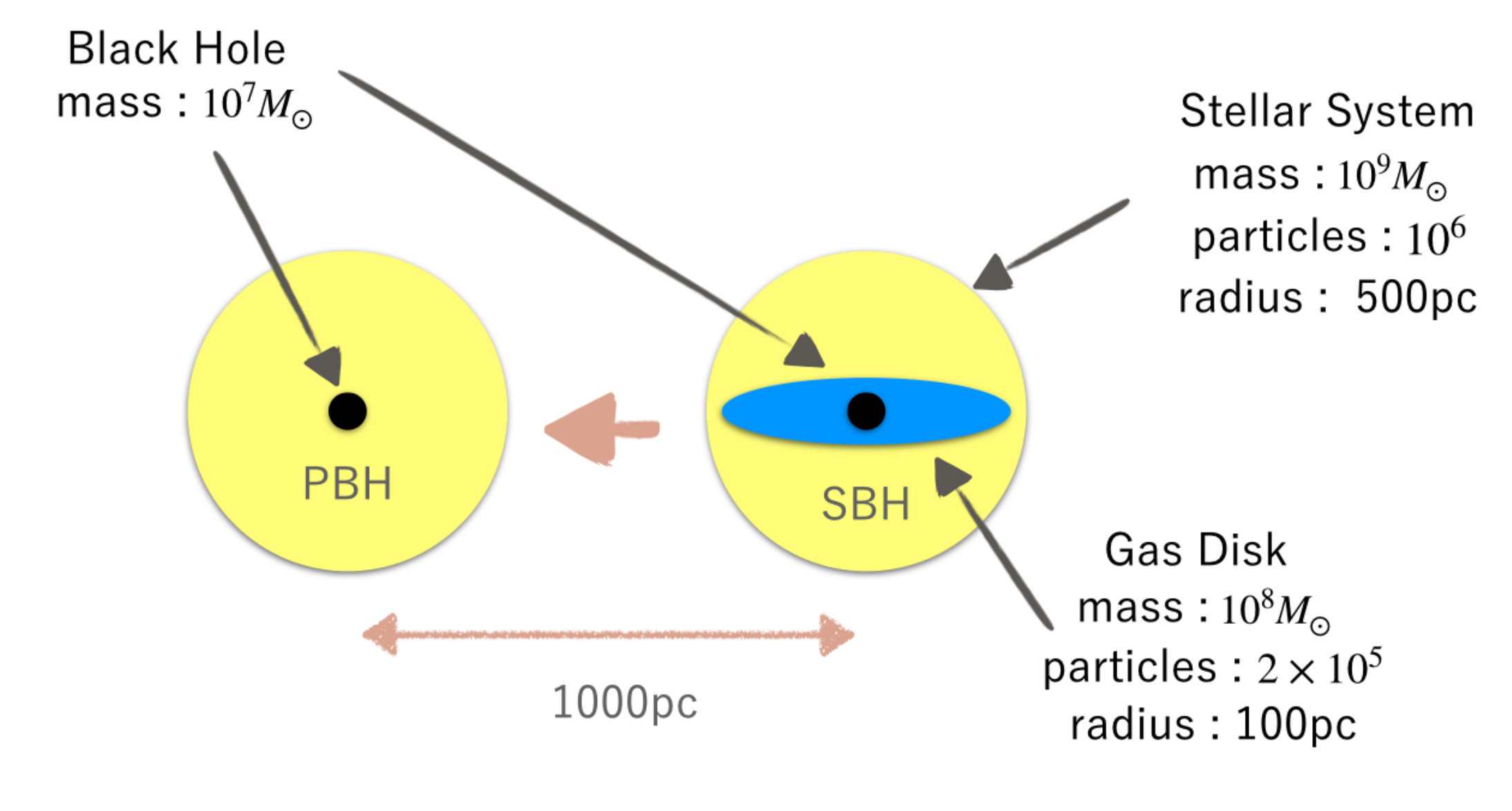}
\caption{Schematic picture of initial setup of the fiducial model (Model hP).}
\label{fig: 1}
\end{figure}

The model setup is schematically shown in Fig. 1.
   A primary black hole ($10^7 M_\odot$) (hereafter, PBH) is surrounded by a spherical stellar system ($10^9 M_\odot$ and a radius of 500 pc)
   represented by $N = 10^6$ particles. Its initial mass profile is a King model ($W_0 = 3$).
A merging system (hereafter, an SBH system) consists of a secondary black hole (SBH, $10^7 M_\odot$) with a rotationally supported gas disk ($10^8 M_\odot$ and a radius of $100$ pc)
represented by $2\times 10^5$ SPH particles, embedded in the same 
spherical stellar system as the PBH system. {The mass resolutions are $10^3 M_\odot$ for stars and $500 M_\odot$ for gas.
The gravitational softening radii are 0.5 pc for both SPH and star particles.}

{Observations suggest that the BH mass ($M_{BH}$) is approximately 0.1\% of the bulge mass of the host galaxies \citep{marconi2003}. 
We do not investigate the entire bulges, but rather focus on their centers; therefore, $M_{BH}$ is assumed to be 1\% of the 
stellar system.}
This SBH system falls toward the PBH system, and they merge.
We follow the fate of the gas supplied by the SBH system. 
{Initially, the PBH system does not include gas. This is because we do not know the exact initial distribution of the gas surrounding SMBHs in general.
In this experiment, we simply follow the fate of the gas supplied by the secondary system. As described below, we also explore multiple mergers, where 
a third system merges into the remnant gas of the previous merger. We find that this second merger
 is not essentially different from the first ``half-dry'' merger.}

We test nine models with different initial positions and angular momenta, {relative to the center of the PBH system}, 
as summarized in Table \ref{table1}.

\begin{table}[h]
    \caption{{Model parameters. $a$: {initial position (x, y, z) of the SBH system (and TBH for Model Multiple) relative to the position of PBH.} $b$: {initial velocity of the SBH system relative to the velocity of PBH.} $c$: energy conversion efficiency for $\dot{M}c^2$ at $r = 1$ pc.}}
  \begin{tabular}{l l l l } 
    \rm Model & \rm $^a$Position [kpc] & \rm $^b$Velocity [pc Myr$^{-1}$] & $^c$AGN Feedback \rm  \\ \hline
     \begin{minipage}{40mm}
      hP (fiducial model)
    \end{minipage} &
    \begin{minipage}{40mm}
      (1, 0, 0)
    \end{minipage} &
    \begin{minipage}{40mm}
      (0, +50.0, 0)
    \end{minipage} &
    \begin{minipage}{40mm}
      $\epsilon= 0.$
    \end{minipage} \\ 

    \begin{minipage}{40mm}
      hPLowAM
    \end{minipage} &
    \begin{minipage}{40mm}
      (1, 0, 0)
    \end{minipage} &
    \begin{minipage}{40mm}
      (0, +12.5, 0)
    \end{minipage} &
    \begin{minipage}{40mm}
          $\epsilon= 0.$
    \end{minipage} \\ 

     \begin{minipage}{40mm}
      hPHighAM
    \end{minipage} &
    \begin{minipage}{40mm}
      (1, 0, 0)
    \end{minipage} &
    \begin{minipage}{40mm}
     { (+25.0, +75.0, 0)}
    \end{minipage} &
    \begin{minipage}{40mm}
        $\epsilon= 0.$
    \end{minipage} \\ 

     \begin{minipage}{40mm}
      hR
    \end{minipage} &
    \begin{minipage}{40mm}
      (1, 0, 0)
    \end{minipage} &
    \begin{minipage}{40mm}
      (-6.25, -50.0, 0)
    \end{minipage} &
    \begin{minipage}{40mm}
          $\epsilon= 0.$
    \end{minipage} \\ 

     \begin{minipage}{40mm}
      sP
    \end{minipage} &
    \begin{minipage}{40mm}
      (1, 0, 1)
    \end{minipage} &
    \begin{minipage}{40mm}
      (-25.0, +50.0, -25.0)
    \end{minipage} &
    \begin{minipage}{40mm}
          $\epsilon= 0.$
    \end{minipage} \\ 

     \begin{minipage}{40mm}
      sPLowAM
    \end{minipage} &
    \begin{minipage}{40mm}
      (1, 0, 1)
    \end{minipage} &
    \begin{minipage}{40mm}
      (-6.25, +50.0, -6.25)
    \end{minipage} &
    \begin{minipage}{40mm}
      $\epsilon= 0.$
    \end{minipage} \\ 

     \begin{minipage}{40mm}
      AGN
    \end{minipage} &
    \begin{minipage}{40mm}
      (1, 0, 0)
    \end{minipage} &
    \begin{minipage}{40mm}
      (-6.25, +50.0, 0.0)
    \end{minipage} &
    \begin{minipage}{40mm}
            $\epsilon= 2\times 10^{-3}$
    \end{minipage} \\ 

     \begin{minipage}{40mm}
      LowAGN
    \end{minipage} &
    \begin{minipage}{40mm}
      (1, 0, 0)
    \end{minipage} &
    \begin{minipage}{40mm}
      (-6.25, +50.0, 0.0)
    \end{minipage} &
    \begin{minipage}{40mm}
    $\epsilon= 2\times 10^{-4}$
    \end{minipage} \\ 

     \begin{minipage}{40mm}
      HighAGN
    \end{minipage} &
    \begin{minipage}{40mm}
      (1, 0, 0)
    \end{minipage} &
    \begin{minipage}{40mm}
      (-6.25, +50.0, 0.0)
    \end{minipage} &
    \begin{minipage}{40mm}
   $\epsilon= 2\times 10^{-2}$
    \end{minipage} \\
    
         \hline
     \begin{minipage}{40mm}
     {Multiple (SBH)}
    \end{minipage} &
    \begin{minipage}{40mm}
      (1, 1, 0)
    \end{minipage} &
    \begin{minipage}{40mm}
      (-6.25, +50.0, -6.25)
    \end{minipage} &
    \begin{minipage}{40mm}
   $\epsilon= 0.$
    \end{minipage} \\
         \begin{minipage}{40mm}
     {Multiple (TBH)}
    \end{minipage} &
    \begin{minipage}{40mm}
      (1, 0, 1)
    \end{minipage} &
    \begin{minipage}{40mm}
      (-6.25, +50.0, -6.25)
    \end{minipage} &
    \begin{minipage}{40mm}
   $\epsilon= 0.$
    \end{minipage} \\    
    
    \hline  \label{table1} %
  \end{tabular}
  \end{table}

{
In reality, galaxies are formed through multiple merger processes, and 
the mass of the central region of galaxies increases in a stepwise manner \citep{saitoh2004}.
It is not clear whether these multiple mergers enhance
mass accretion to the SMBHs.
To model this multiple merger, we investigate a merger process with a third BH system ({Model Multiple}). {The third BH (TBH) system 
has the same structure as the SBH system, and its initial location and velocity are also shown in Table 1.}
%


\subsection{Numerical Methods}   \label{subsec:methods}
%
We use the parallel tree N-body/SPH code ASURA \citep{saitoh2008, saitoh2009, saitoh2013}. Radiative cooling is implemented by a cooling table between $10$K $\sim 10^9$K \citep{wada2009}.
The heating sources include the far ultraviolet radiation (10 $G_0$, where $G_0$ is the galactic local value) and photoelectric heating. The star formation and supernova feedback are based on \citet{saitoh2008}.
Dense and cold SPH particles are probabilistically converted to star particles by assuming Schmidt's law \citep{schmidt1959} with an efficiency 
{relative to the gas mass per free fall time} of 0.0033.
The conditions of star formation are determined by density and temperature thresholds: $n_{th} > 10^3$ cm$^{-3}$ and $T_{th} < 100$ K, respectively.
Supernova feedback is implemented probabilistically based on \citet{okamoto2008}.
A Type-I\hspace{-.1em}I supernova is assumed, and each explosion provides $10^{51}$ ergs into 32 surrounding SPH particles as a form of thermal energy.

The mass accretion rate to the BHs (i.e., PBH, SBH, and TBH) is determined by
the following conditions for SPH particles
inside an accretion radius $r_{acc}$:
1) their kinetic energy is lower than the gravitational energy, and 2) the angular momentum of an SPH particle $J_i$ is lower than a threshold:
$
J_i < J_{acc}  = r_{acc} \sqrt{{GM_{BH}}/(r_{acc}^2+r_0^2)^{1/2}},
$
where {$r_0$ is a gravitational softening of BHs}. Here, we assume that $r_0 = 0.5$ pc and {$r_{acc} = 1.0$ pc}.

{AGN feedback energy $\Delta E$ is kernel-weighted and distributed to 32 SPH particles around the BH particles in the form of thermal energy \citep[e.g.,][]{Springel2005},
which is then calculated using the gas mass accretion rate $\dot{M} $ at $r = r_{acc}$, i.e.,
$\Delta E =  \epsilon \dot{M}c^2$.}
where $\epsilon$ is the energy conversion efficiency. 
We explore three AGN models with $\epsilon = 2\times 10^{-2}$ (Model HighAGN),
$\epsilon = 2\times 10^{-3}$ (Model AGN), and $\epsilon = 2\times 10^{-4}$ (Model LowAGN).

%
\section{Results} \label{sec:results}
%
\subsection{Gas accretion to BHs}
PBHs and SBHs approach one another as a result of dynamic friction with the stellar component. This distorts the gas component and forms spiral arms.
{Figure \ref{fig: 1evolve} shows the evolution of the density distributions in the central 2 kpc $\times$ 2 kpc and 0.4 kpc $\times$ 0.4 kpc in Model hP. }
When the two BH systems merge, tidal spiral arms are formed ($t =$ 20 Myr and 25 Myr), and the system settles to a smoothed disk ($t =$ 75 Myr).
In this model, the SBH system falls toward the PBH on the $z=0$ plane, and the gas forms a thick disk with $\sim $ 100 pc thickness.
Figure \ref{fig: 2_4model} compares gas density distributions in {the three models (hP, hPLowAM, and sP)} when the SBH and the PBH form a binary with a $\sim$ 50 pc separation. The tidal spiral is not prominent in the model with the retrograde orbit (Model hR).


Figure \ref{fig: 2} shows the evolution of the distance between the PBH and the SBH in six models.
  In Model hR and Model hPLowAM, the SBH approaches the PBH within 10 pc in the first $t \sim 20$ Myr,
 forming a binary with a separation of several parsecs. 
{Note that the distance between the PBH and the SBH shows a complicated behavior 
in Model hP: a sudden increase at $t \sim 50$ Myr and again at $t\sim 80$ Myr. 
This is also seen in 
 Model sPLowAM after $t \sim 70$ Myr. As shown in Fig. \ref{fig: 3} below, the accretion rate to the BHs suddenly increases
 at these times. This is because the BHs gain their angular momentum by 
 accreting the gas; as a result, the orbits of BHs become larger and more elongated. 
 If this is the case, BHs could lose their angular momenta through the dynamic
 friction. Then, the distance between PBH and SBH changes in a nonlinear manner.}
Models with different initial angular momenta and orbits (Models hPHighAM, sP, and sPLowAM) show that the
PBH and the SBH approach at $t \sim 50$ or 100 Myr, but their final distances are similar to those in the other models.

\begin{figure}[h]
\centering
\includegraphics[width = 13cm]{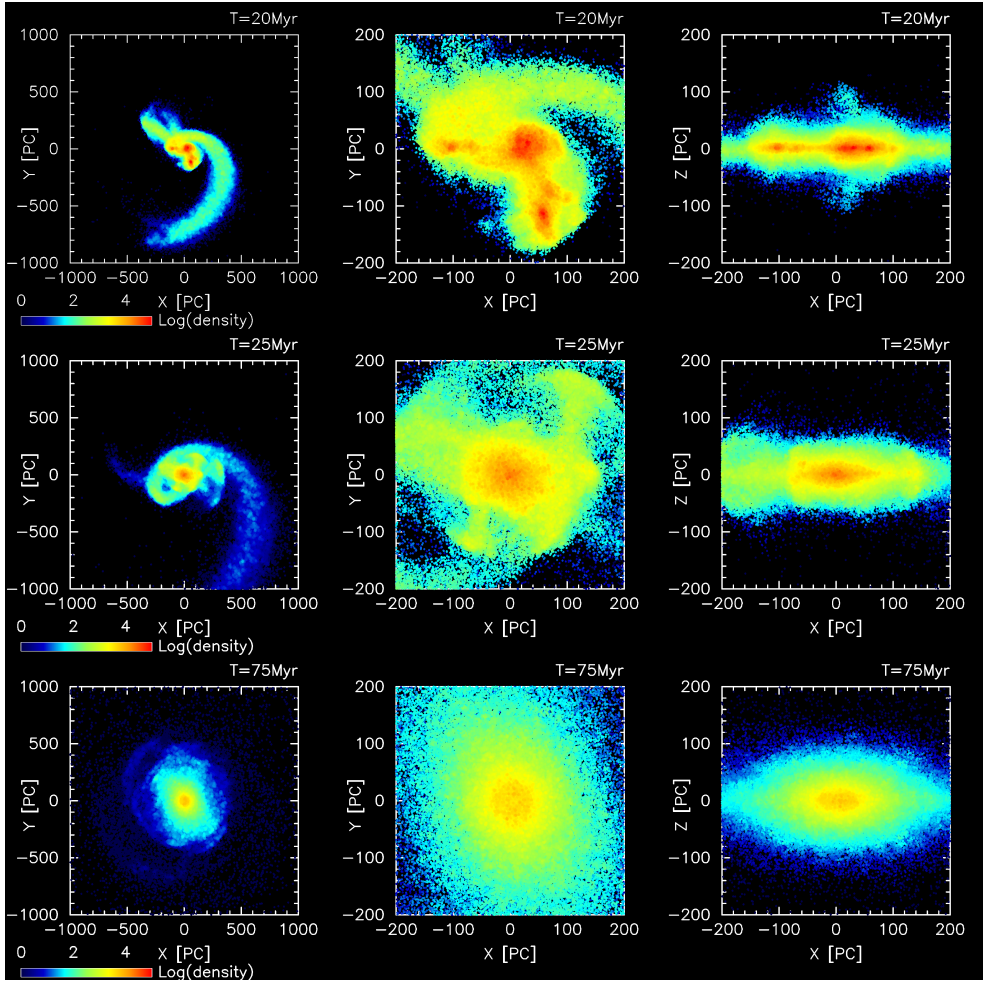}
\caption{{Evolution of gas density in the fiducial model (Model hP). 
Three snapshots, at $t = 20, 25$, and 75 Myr, are shown. 
The left panels show distribution projected onto the $x$-$y$ plane (2 kpc $\times$ 2 kpc), and 
the right two rows are close-ups for the $x$-$y$ and $x$-$z$ planes (0.4 kpc $\times$ 0.4 kpc).
The axis units are in parsecs. The color bar represents log-scaled density ($M_\odot$ pc$^{-3}$).} }
\label{fig: 1evolve}
\end{figure}
\begin{figure}[h]
\centering
\includegraphics[width = 10cm]{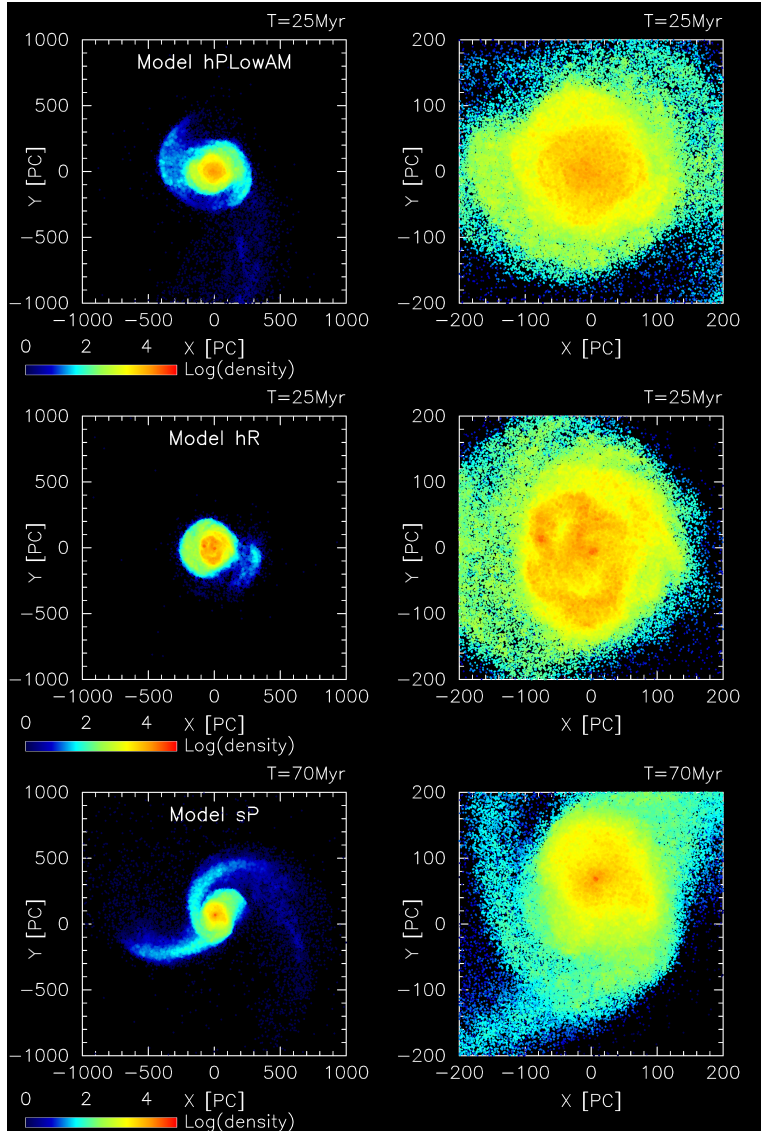}
\caption{{Comparison of the gas density distributions in three models (hPLowAM at $t=$ 25 Myr, hR at $t=$ 25 Myr, and sP at $t = 70$ Myr). Here, the PBH and SBH are separated by approximately 50 pc in Models hPLowAM and sP, and 100 pc in Model hR. 
Two arrows in the top left panel indicate the positions of black holes (PBH and SBH) in Model hP. The axis unit is in parsecs. The color represents log-scaled density ($M_\odot$ pc$^{-3}$).} }
\label{fig: 2_4model}
\end{figure}

\begin{figure}[h]
\centering
\includegraphics[width = 18cm]{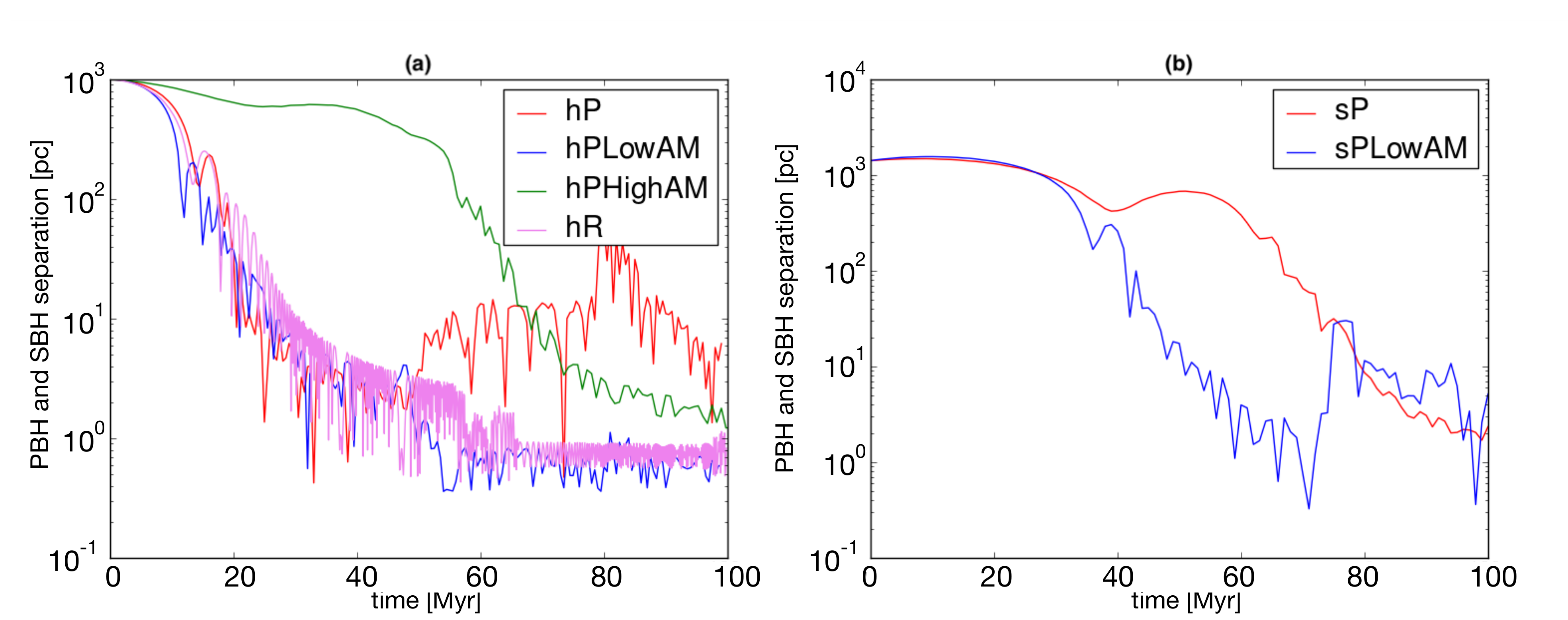}
\caption{{Time evolution of the distance between PBH and SBH in six models. (a) Models hP, hPLowAM, hPHighAM, and hR. (b) Models sP and sPLowAM, {in which the orbit of the SBH system} is not in the same plane as the gas disk around SBH.}}
\label{fig: 2}
\end{figure}

\begin{figure}[h]
\centering
\includegraphics[width = 18cm]{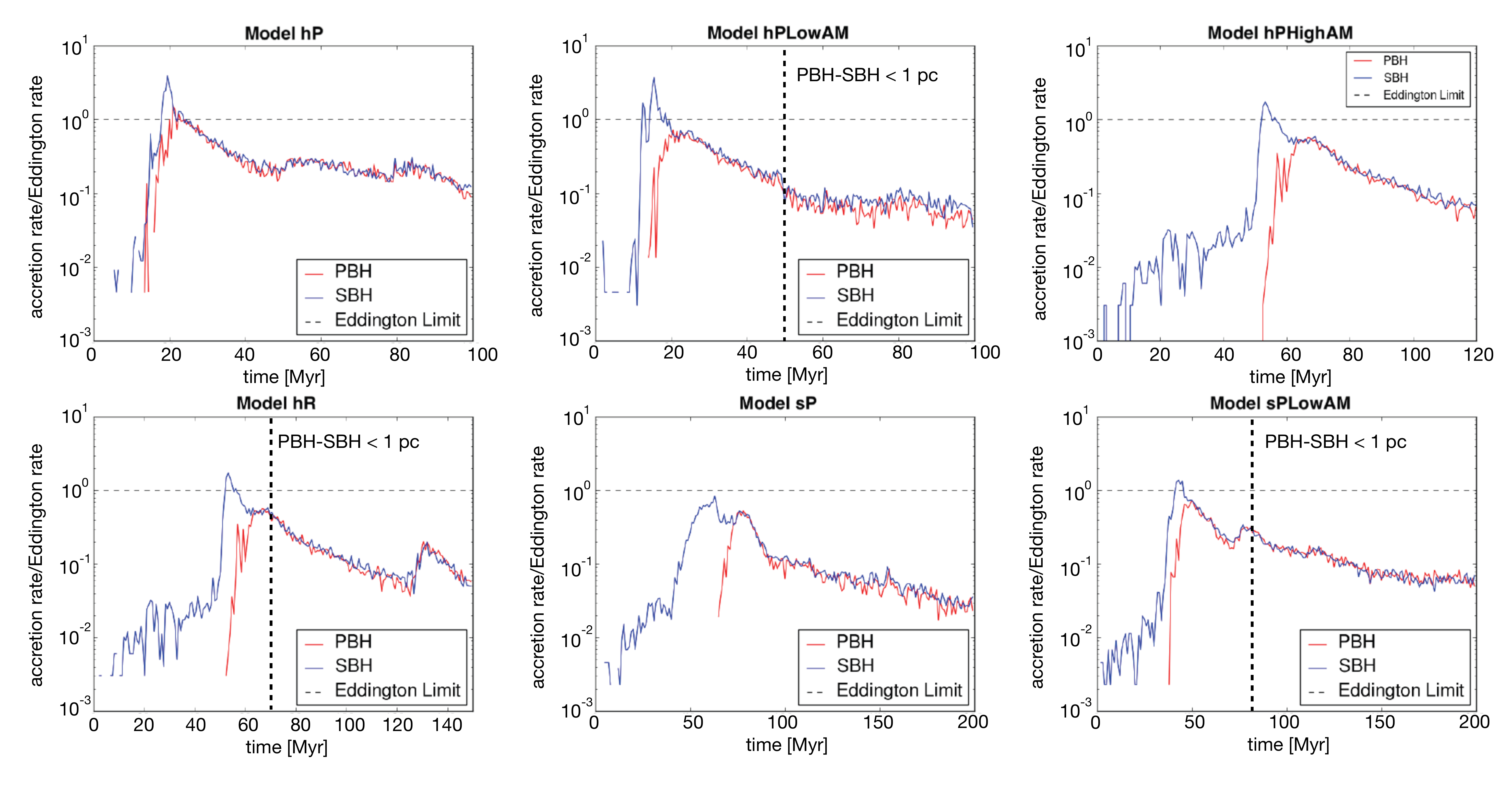}
\caption{{Evolution of the mass accretion rates normalized by the Eddington rate of PBH and SBH in six models. In three models (hPLowAM, hR, and sPLowAM), the separation between the PBH and SBH becomes smaller than 1 pc, which is the radius where the accretion rate is measured, after $t  \sim 70$ Myr (see Fig. \ref{fig: 2}). The time is shown by the vertical dashed lines. }}
\label{fig: 3}
\end{figure}

The mass accretion rates of the PBH and the SBH as a function of time in the six models in Fig. \ref{fig: 2} are plotted in Fig. \ref{fig: 3}.
All models show a rapid mass accretion phase, which lasts $\sim 10$ Myr, followed by
a quasi-steady accretion phase, where
the accretion rate gradually decreases toward 10\% or less of the maximum accretion rate.
{The peak accretion rates of the SBH (e.g., $t \sim 20$ Myr in Model hP) are 2--4 times higher than those of the PBH.
This is primarily because the gas around the SBH is tidally disturbed as the two BHs approach each other
and because the PBH is initially not surrounded by the gas (Fig. 1).
However, when the third BH system (TBH) merges into the PBH-SBH system (Model Multiple),
 both the accretion rates for the PBH and the SBH are slightly smaller than those for TBH (see \S 3.4 and Fig. \ref{fig: 18}).}

\begin{figure}[h]
\centering
\includegraphics[width = 18cm]{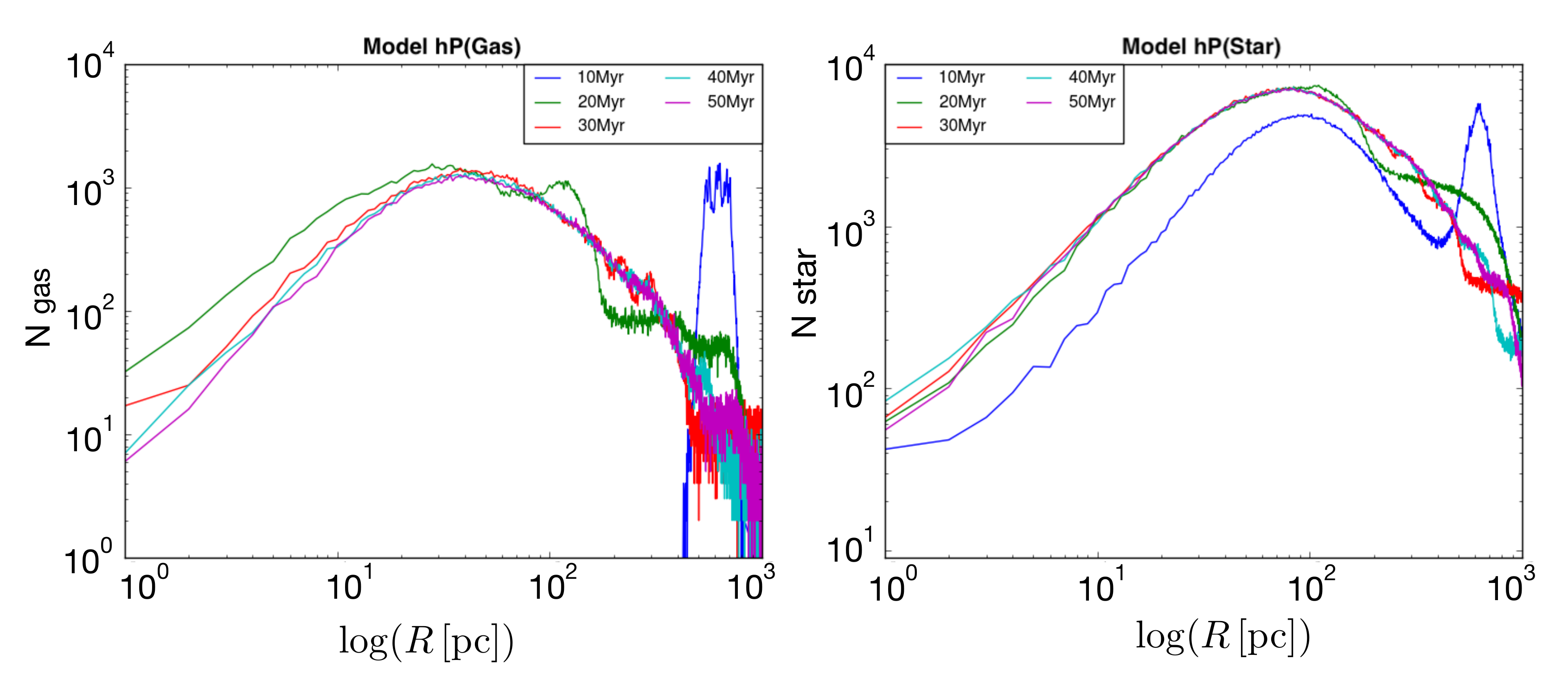}
\caption{ {(left) Evolution of the radial profiles of the number of SPH particles per bin in the fiducial model (Model hP). (right)  Same as the left panel, but for stars.}}
\label{fig: 7}
\end{figure}

\begin{figure}[h]
\centering
\includegraphics[width = 9cm]{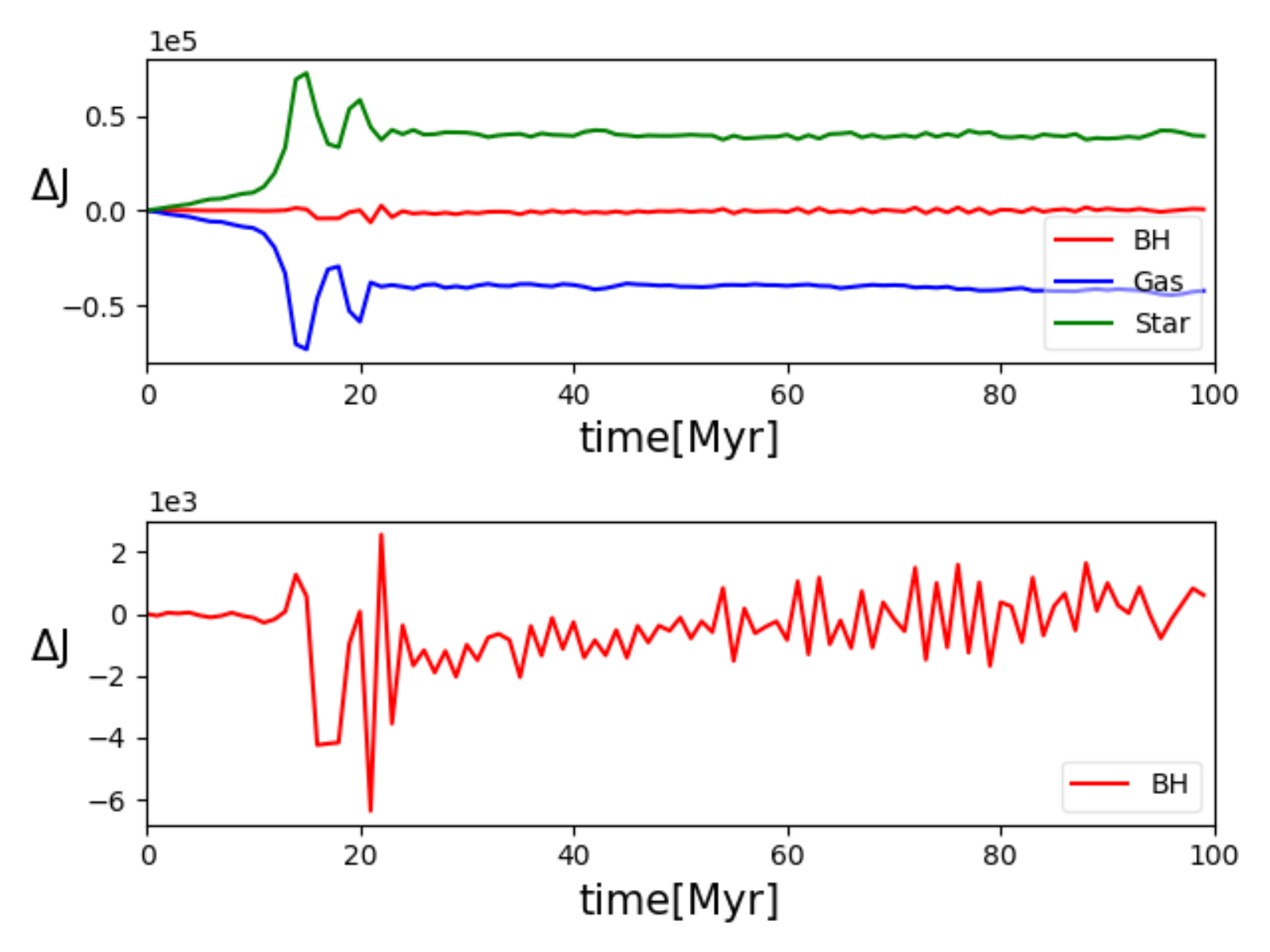}
\caption{(top) {Changes in the total angular momentum (the reference frame is the coordinate origin) for the $z$-direction from those in the initial values} ($\Delta J$) for PBH$+$SBH, gas, and stars in 
Model hP.  (bottom)   $\Delta J$ only for the BH system.}

\label{fig: 4}
\end{figure}

{
 In the fiducial model (Model hP), rapid accretion occurs as the two black holes approach
($t\sim 20$ Myr). Their maximum mass accretion rates to the PBH and the SBH are $0.3  M_\odot$ yr$^{-1} \simeq 1.5 \dot{M}_{Edd}$
and $0.9  M_\odot$ yr$^{-1} \simeq 4.5 \dot{M}_{Edd}$, where $\dot{M}_{Edd} \equiv 4\pi G M_{PBH} m_p/(\epsilon \sigma_T c)$ with
 the proton mass $m_p$, the Thomson scattering cross-section $\sigma_T$, and the efficiency $\epsilon = 0.1$.
The super-Eddington accretion for the SBH lasts over 10 Myr.
Models hPLowAM, hPHighAM, sP, and sPLowAM show similar behavior, 
but the maximum accretion rate to the PBH is sub-Eddington ($\sim$ 0.8 $\dot{M}_{Edd}$).
}
In the fiducial model (Model hP), which has
a higher initial angular momentum than Model hPLowAM, {the mass accretion rate in the quasi-steady accretion phase ($t =40$ Myr $\sim$ 100 Myr)
 is larger than that in Model hPLowAM.}
The accretion rates in Model sP, which has an inclined orbital plane for the SBH, do not exceed the Eddington limit for either the PBH or the SBH.
This is not the case if the initial angular momentum is low (Model sPLowAM), where the accretion rate for the SBH slightly exceeds the
Eddington limit for a short period\footnote{
{Note that in models hPLowAM, hR, and sPLowAM, the separation of the PBH and the SBH becomes smaller than 1 pc after 50 Myr. 
Because the accretion rate is measured at $r = 1$ pc from the PBH, the difference in the accretion rates for the PBH and the SBH shown in Fig.  \ref{fig: 3} would be
meaningless. The time is shown by the vertical dashed lines in Fig. \ref{fig: 3}. } }.

Figure \ref{fig: 7} shows the radial distributions of the gas and star particles relative to the position of PBHs and their time evolution in
the fiducial model (Model hP). {Note that the bump at $r \sim 1$ kpc for $t = 10$ Myr corresponds to the SBH system approaching 
the PBH.}
The gas inside several 100 pc for the PBH shows no significant change in its distribution
after $t =$ 30 Myr. The number of gas particles in the central 100 pc
gradually decreases, reflecting mass accretion to the PBH [Fig. \ref{fig: 7} (left)].
The stellar distributions do not change after $t \sim 20$ Myr (i.e., the time when the PBH and the SBH rapidly approach each other).
This implies that there is no significant transfer of angular momentum to stars from the gas and BHs in the late phase
of evolution (see also Fig. \ref{fig: 4}).
%

{In summary, the accretion rates of PBHs and SBHs in both the rapid accretion and quasi-steady phases 
are not significantly affected by the choice of the orbit of the merging system and the 
initial angular momentum. Depending on the initial orbits of the BHs, the time of peak accretion changes. 
The variation of the accretion rate among the models is within a factor of 2--5. 
}

{ Figure \ref{fig: 4}(top) shows the changes in the total angular momenta ($z$-direction) of gas, stars, and BHs from
 their initial values. 
The plot indicates that the angular momenta of the gas and stars change significantly at approximately $t \sim$ 20 Myr and 
are transferred to the stellar component from the gas.
  After this rapid accretion phase, there is no significant transfer of the angular momentum between the gas and stars.
In contrast, the variation of the angular momentum for BHs is approximately 100 times smaller than those for gas and stars (Fig.  \ref{fig: 4}(bottom)).
The change in the angular momentum becomes slightly larger after $t \sim 50$ Myr, which is caused by gas accretion to the BH system.
This is also seen in the change in the orbits of the BHs in Fig. \ref{fig: 2}.}
  
%
\subsection{Two accretion phases}
%
To understand the mechanism behind the angular momentum transfer and the mass accretion shown in \S 3.1,
torques resulting from gravity $\vec{\tau}_G$ and from {the turbulent pressure gradient $\vec{\tau}_{turb}$ and their evolution are examined, where
$\vec{\tau}_G \equiv \vec{r} \times \nabla \Phi_G $ and $\vec{\tau}_{turb} \equiv \vec{r} \times \nabla (\rho \, \sigma_g^2)/\rho $ with 
the velocity dispersion of the gas $\sigma_g$. }
The gravitational potential $\Phi_G$ is calculated in ASURA code using the tree algorithm, including the gas and stellar components.
{Radial distributions of the ratio of the $z$-components of the two torques, i.e., $\tau_{G,z}/\tau_{turb, z}$ at
$t =$ 15, 20, and 75 Myr are plotted for Model hP in Fig. \ref{fig: 5}. We found that the gravitational torque dominates
the turbulent pressure torque in the early phase of merger (i.e., $t = 15$ Myr), but $\tau_{G,z}/\tau_{turb, z}$ decreases with time, 
and the outer part of the circum-binary disk ($ R \gtrsim 50$ pc) is dominated by the turbulent pressure torque (i.e., $\tau_{G,z}/\tau_{turb, z} < 1$)
after the rapid fueling phase.}
%

%

\begin{figure}[h]
\centering
\includegraphics[width = 10cm]{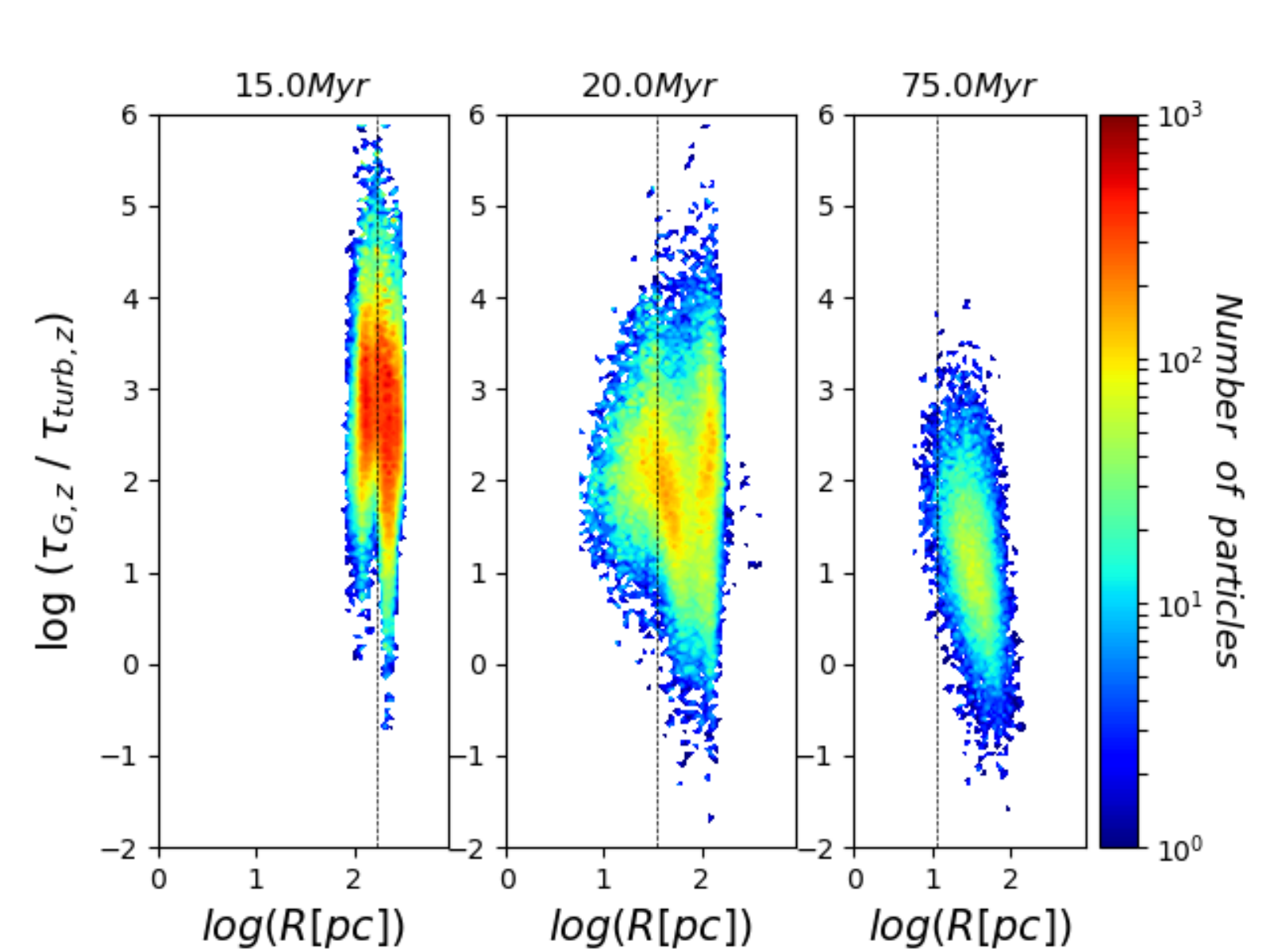} 
\caption{ {Radial distributions of the ratio of the $z$-components of the torques originated in the gravity and
turbulent pressure, i.e., $\tau_{G,z}/\tau_{turb, z}$
for $t=$ 15, 20, and 75 Myr.
$R$ represents distance from PBH. 
The colors represent number of SPH particles for a given torque ratio and the radius bin. 
The positions of SBH are shown by the vertical dotted lines.
}}
\label{fig: 5}
\end{figure}

As mentioned above, for the
fiducial Model hP, the rapid accretion phase is followed by a quasi-steady accretion phase after $t\sim 20$ Myr.
{In this phase,
there is no significant exchange of the total angular momentum between gaseous and stellar components (see Fig. \ref{fig: 4}). 
The variation of the angular momenta of BHs is less than 1\% of those for gas and stars. }
{Figure \ref{fig: 9} shows the time evolution of the turbulent viscous time scale in units of the disk rotation period.}
{The turbulent viscous time scale $\delta t_{vis}$ can be estimated as $ \delta t_{vis} \sim R_{disk}^2/ (\sigma_{z}\,  h_{disk})$\citep{lin_pringle1987},
where $\sigma_{z} $ is the velocity dispersion in the $z$-direction,  $h_{disk}$ is the scale height of the
dense gas disk ($\sim 100$ pc), and $R_{disk}$ is the radius of the gas disk around the PBH. $R_{disk}$ and $h_{disk}$ are determined by 
setting a density threshold ($100 M_\odot $ pc$^{-3}$). The kernel size of the SPH particles is approximately 1 pc for the gas in the disk.}
This shows that the viscous time scale is comparable with the rotational period $t_{rot}$, {which is calculated from 
the mass inside $r = R_{disk}$. After $t \sim 50$ Myr, the ratio $\delta t_{vis}/t_{rot}$ is approximately equal, suggesting that
the angular momentum transfer in the late-phase accretion is mostly determined by the turbulent viscosity in the disk.
{As seen in Fig. \ref{fig: 5} at t = 75 Myr, the torque originating in the turbulent pressure gradient also contributes
to the angular momentum transfer, especially for the outer part of the disk in the quasi-steady accretion phase.}
The decreasing trend of $\delta t_{vis}/t_{rot}$ primarily occurs as
the size of the disk $R_{disk}$ decreases. Because $R_{disk}$ is defined by setting a density threshold, 
it is affected by the tidal structures of the merger remnant, particularly during the early phase of evolution ($t \lesssim 20$ Myr) (see Fig. \ref{fig: 1evolve}).
 The disk size also shrinks as a result of mass accretion.}

\begin{figure}[h]
\centering
\includegraphics[width = 10cm]{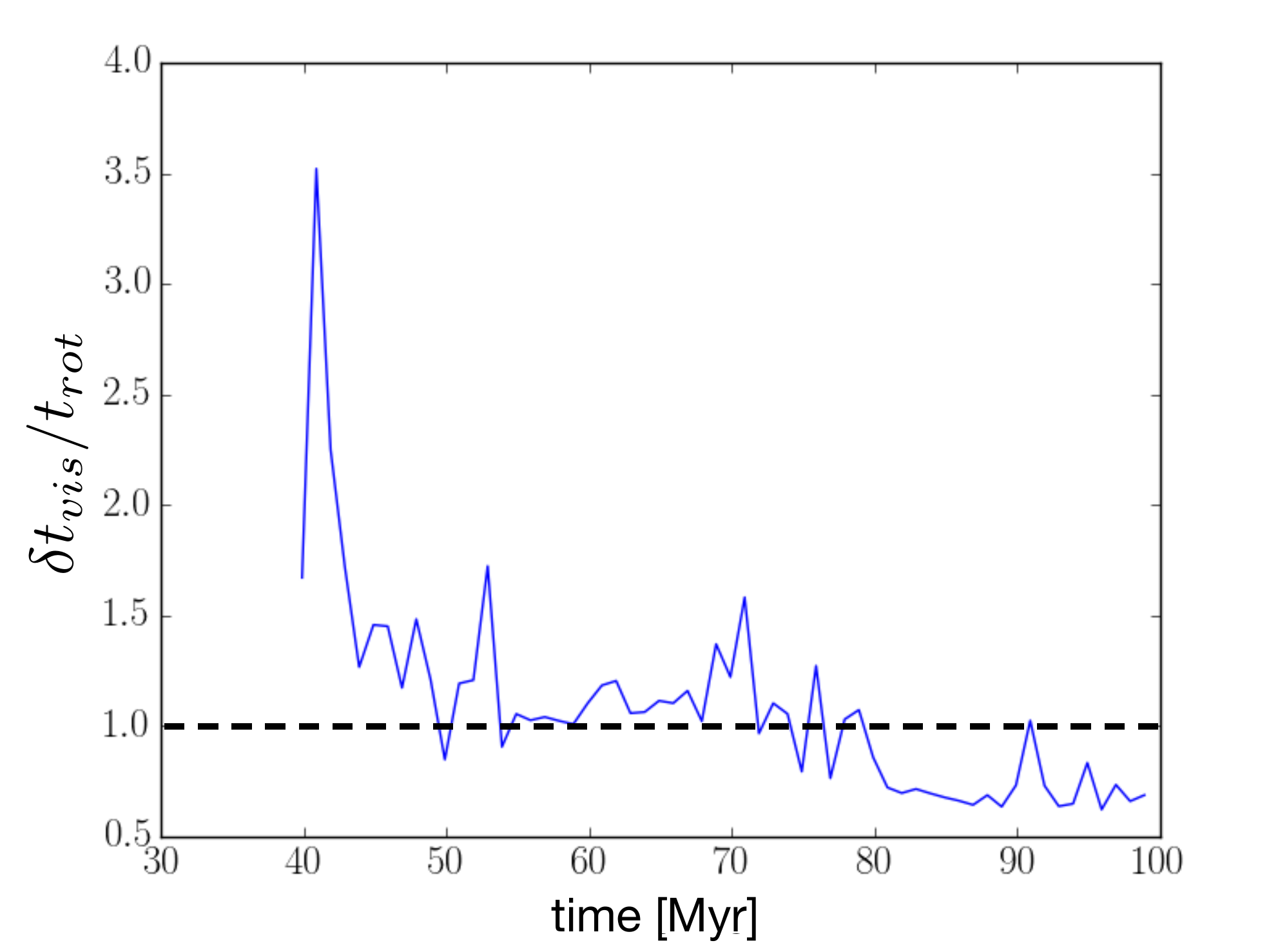}
\caption{{Evolution of viscous time scale ($\delta t_{vis}$) resulting from turbulent viscosity, normalized by
rotational time scale $t_{rot}$ after the quasi-steady circumnuclear disk is formed (i.e., $t > 40$ Myr). } }
\label{fig: 9}
\end{figure}

\subsection{Effect of feedback on accretion}

We examine how the energy feedback from AGNs affects the mass accretion processes.
{Figure \ref{fig: feedback} is a close-up of the density field in 
the central 200 pc $\times$ 200 pc for Model hP (without AGN feedback) 
and two models with AGN feedback.
 We found that the AGN feedback does not change the large-scale morphology (cf. Fig. 2), but it 
makes the gas diffuse around the
BHs. In Model HighAGN, where the feedback energy rate is calculated as $0.02 \dot{M} c^2$,
at a mass accretion rate to $r = $ 1 pc, $\dot{M}$,
almost no high-density gas remains around the BHs.
This is in contrast to Model LowAGN, where the energy conversion efficiency is 1/100 of that in Model HighAGN.}

Figure \ref{fig: 14} compares the time evolution of the mass accretion rates {to SBH} in the three feedback models (Models AGN,
HighAGN, and LowAGN) with that of the fiducial model (Model hP).
{
There is no significant difference between the mass accretion rates of
Model hP and Model LowAGN, and the accretion rate in Model AGN is slightly smaller than those in the other two models after $t \sim 50$ Myr.}
All three models show a peak of mass accretion, {$\dot{M} \simeq 3-4 M_\odot$ yr$^{-1}$} at approximately $t \sim 20$ Myr.
{However, the mass accretion rate at the peak is sub-Eddington ($0.7 M_\odot$ yr$^{-1}$) in model HighAGN, and
rapidly decreases to $\sim $ 1/10  in Model hP and Model LowAGN in the late accretion phase ($t > 20$ Myr).
In Model HighAGN, 2\% of $\dot{M} c^2$ at $r = $1 pc is supplied to the
the circumnuclear gas. This result shows that the conversion efficiency, ``2\%,'' is too large to contain the mass accretion to SMBHs.}
In other words, AGN feedback and black hole growth (i.e., mass accretion) can coexist in galaxy merger simulations if the feedback efficiency at 1 pc is $\sim 0.02-0.2\%$.
Even with this feedback, BHs with $10^7 M_\odot$ in our models double their mass in 100 Myrs for a gas supply of $10^8 M_\odot$.

\begin{figure}[h]
\centering
\includegraphics[width = 18cm]{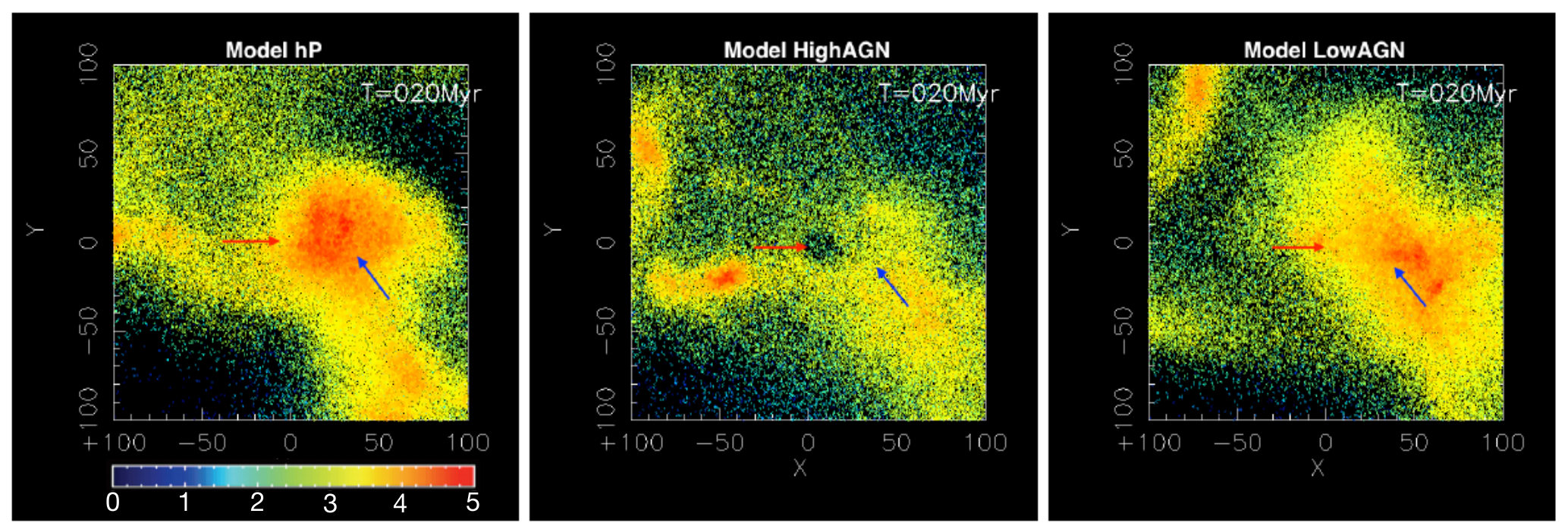}
\caption{Gas density distributions of central 100-pc regions in Models hP, HighAGN, and LowAGN at $t = 20$ Myr. Positions of PBHs and SBHs 
are shown by red and blue arrows. {The color bar represents 
the log-scaled gas density in $M_\odot$ pc$^{-3}$. }　}
\label{fig: feedback}
\end{figure}
%

\begin{figure}[h]
\centering
\includegraphics[width = 11cm]{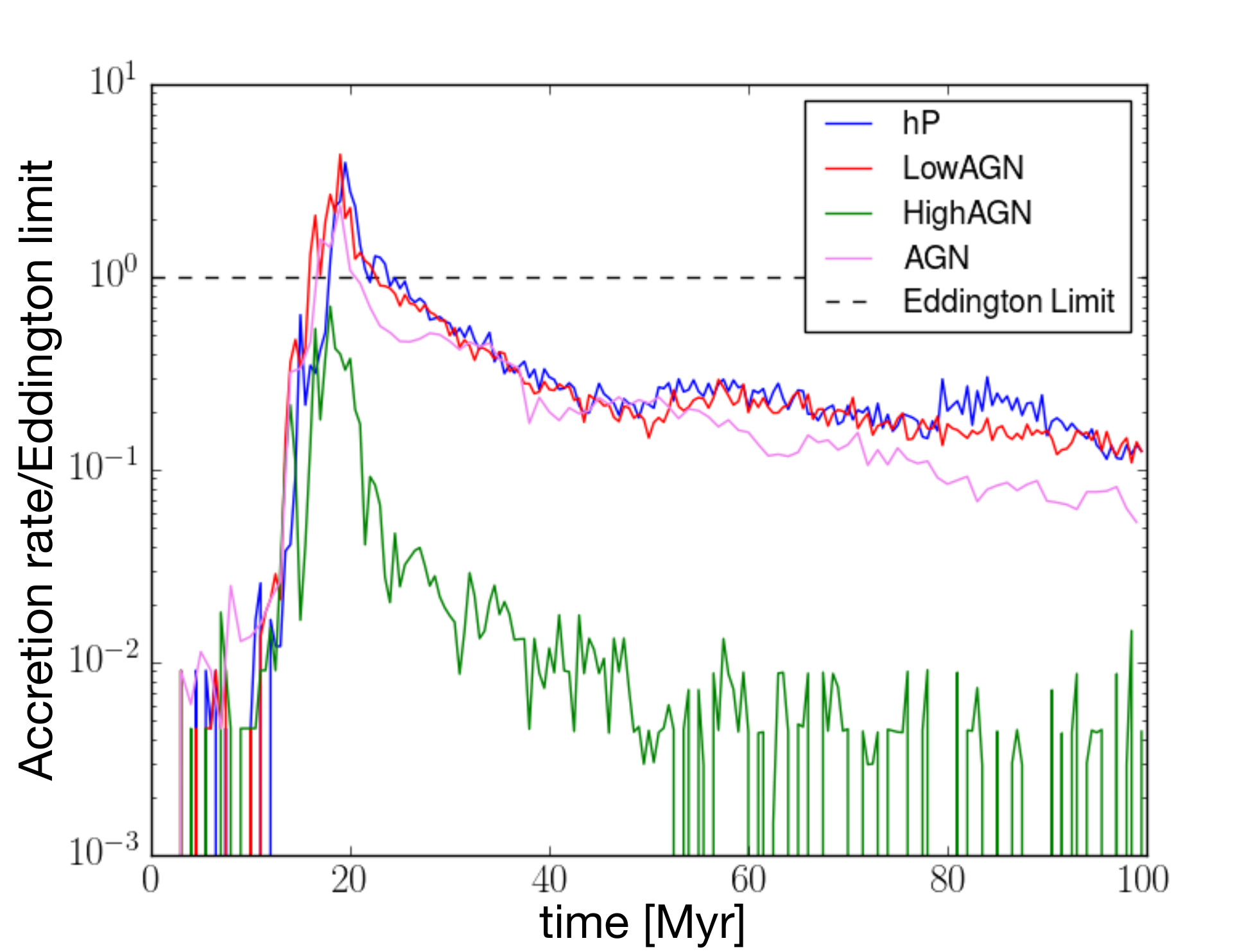}
\caption{Mass accretion rates {for SBH} normalized by Eddington rate for Models hP, LowAGN, HighAGN, and AGN.}
\label{fig: 14}
\end{figure}

\subsection{Effects of multiple mergers}

During galaxy formation, galactic mergers can be episodic.
Major or minor mergers should be more frequent at
high-z, and this can enhance the mass accretion to SMBHs.
{To explore this in our numerical experiments,
 we conducted a ``secondary'' merger after the first merger between PBH and SBH systems ({Model Multiple, see Table 1}).} 
  Figure \ref{fig: 18} shows the evolution of the bolometric luminosity of PBHs, SBHs, and the secondary added BH system (TBH).
  {The bolometric luminosity is assumed to be $L_{AGN} =  0.1 \dot{M} c^2$, where $\dot{M}$ is an accretion rate to PBHs, SBHs, or
  TBHs within 1 pc from each BH.}
This model does not include the AGN feedback, but the luminosity should not be significantly different if the efficiency is 0.2\% or lower, as
inferred from Fig. \ref{fig: 14}.
The luminosity of the {SBH} becomes super-Eddington for a short period ($ <1$ Myr) around the first merger, but
quickly reduces to sub-Eddington.
The second merger occurs and the rapid accretion appears again at $t\sim 90 $ Myr, but
the luminosity does not exceed the Eddington luminosity.
The luminosities of the three BHs are almost comparable after the rapid accretion phases ($t > 100$ Myr), at $\sim$ 10\% of the Eddington luminosity.
{
This result suggests that the amount of accreted gas mostly depends on the total gas mass 
supplied by the merging systems (in our case, SBH or TBH), 
and the mass accretion at each event may not be significantly enhanced by 
episodic mergers. 
However, this is inconclusive, considering 
the present ideal model set-up. It should be verified by future high-resolution, cosmological simulations.}

\begin{figure}[h]
\centering
\includegraphics[width = 10cm]{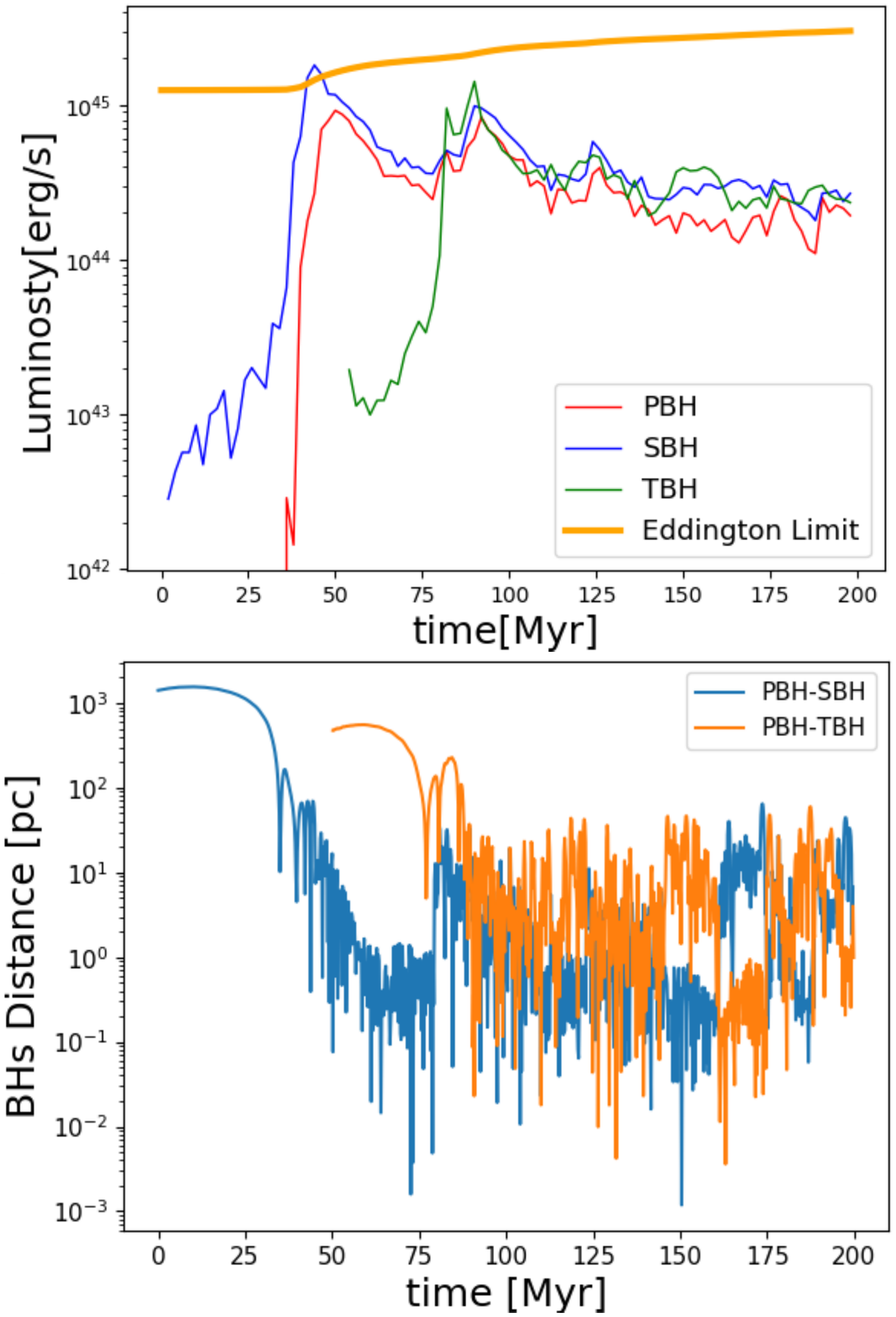}
\caption{(top) Luminosities resulting from mass accretion to three BHs, i.e., PBH, SBH, and TBH, plotted
as a function of time. Eddington luminosity {for PBH} is also plotted as a yellow solid line. 
{(bottom)  Time evolution of relative distance between PBH and SBH (orange), and PBH-SBH (blue). } }
\label{fig: 18}
\end{figure}

%
\section{Discussion: Origin of obscured AGNs}
%

AGNs are often buried in dense gas in ultraluminous infrared galaxies (ULIRGs). Observations suggest that the fraction of
Compton-thick AGNs in late-merger galaxies is higher than that in ULIRGs \citep{ramos-almeida2017, ricci2017}. \citet{buchner2015} also showed that 
Compton-thick AGNs account for approximately 40\% of the $\sim$ 2000 observed AGNs.
Recent wide-field AGN optical surveys such as the Subaru Hyper Sprime-Cam, 
Wide-field Infrared Survey Explorer (WISE), and Sloan Digital Sky Survey (SDSS) have revealed that 
galaxies undergoing mergers are several times more likely to contain luminous obscured
AGNs than non-interacting galaxies \citep{weston2017, goulding2018b}.
However, the origin of the obscuration is not clear. For example, the large column density can be caused either by dense circumnuclear gas or more diffuse ISM on a much larger scale.

To see how the AGN activity is obscured in our models, we 
show the evolution of the column density and the star formation rate of Model AGN in Fig. \ref{fig: 19}.
{The column densities are obtained by averaging the gas density along 
two lines of sight toward PBHs, i.e., edge-on (in the $x-y$ plane) and face-on (in the $x-z$ plane).}
There is some dependence on the line-of-sight directions.
For the edge-on, i.e., the direction in which the secondary system is accreting, the column density exceeds $N = 10^{24}$ cm$^{-2}$
for several Myrs, and it gradually decreases in the following several tens of Myrs.
For face-on, the nucleus does not become Compton-thick, even in the rapid accretion phase.
The column density for {face-on} is a factor of approximately 2 lower than that for edge-on.
{The nucleus is obscured with $N > 10^{24}$ cm$^{-2}$ around $t = 20$ Myr for the edge-on view.
This is mostly caused by the circumnuclear gas at $r < 100$ pc from the PBH. 
This result is consistent with recent numerical results reported by \citet{trebitsch2019}, 
in which cosmological radiative hydrodynamic simulations of galaxy formation 
are used to study the nuclear obscuration.}

The star formation rate (SFR) {around PBHs, which is mostly caused at $r \lesssim 100$ pc} is also plotted in Fig. \ref{fig: 19}.
SFR rapidly increases during the merger, but it quickly decreases after $\sim $ several Myrs.
{This active star-forming phase appears at approximately $t \sim 18$ Myr, which corresponds to the phase of increasing column density}
\footnote{{The peak SFR, $\sim 0.25 M_\odot$ yr$^{-1}$, is approximately ten times larger than 
the value expected for the gas density within 100 pc from PBHs, i.e., $\Sigma_g \sim 500 M_\odot$ pc$^{-2}$, assuming
the SFR-gas surface density relation in starburst galaxies \citep[e.g.,][]{bigiel2008}.}}.

We also found that the column density in Model Multiple shows no clear dependence on 
the line-of-sight angles after the second merger occurs. As expected, the nucleus should be more or less spherically 
obscured by many merger events from random directions if outflows caused by the AGN activities do not significantly affect 
the gas distribution. 

We plot the evolution of the bolometric luminosity and the column density of Model AGN in Fig. \ref{fig: 20}. The time evolution is
represented by the color of each circle (edge-on) or triangle (face-on).
We find that the bright AGN phase ({$L_{bol} > 10^{45}$ erg s$^{-1}$})
is obscured by a large column density ($N > 10^{24}$ cm$^{-2}$ for the edge-on view.
This ``obscured AGN'' phase ends after 30 Myrs, with $N < 10^{24}$ cm$^{-2}$ and $L_{bol} \simeq  10^{44}$ erg s$^{-1}$.

\begin{figure}[h]
\centering
\includegraphics[width = 12cm]{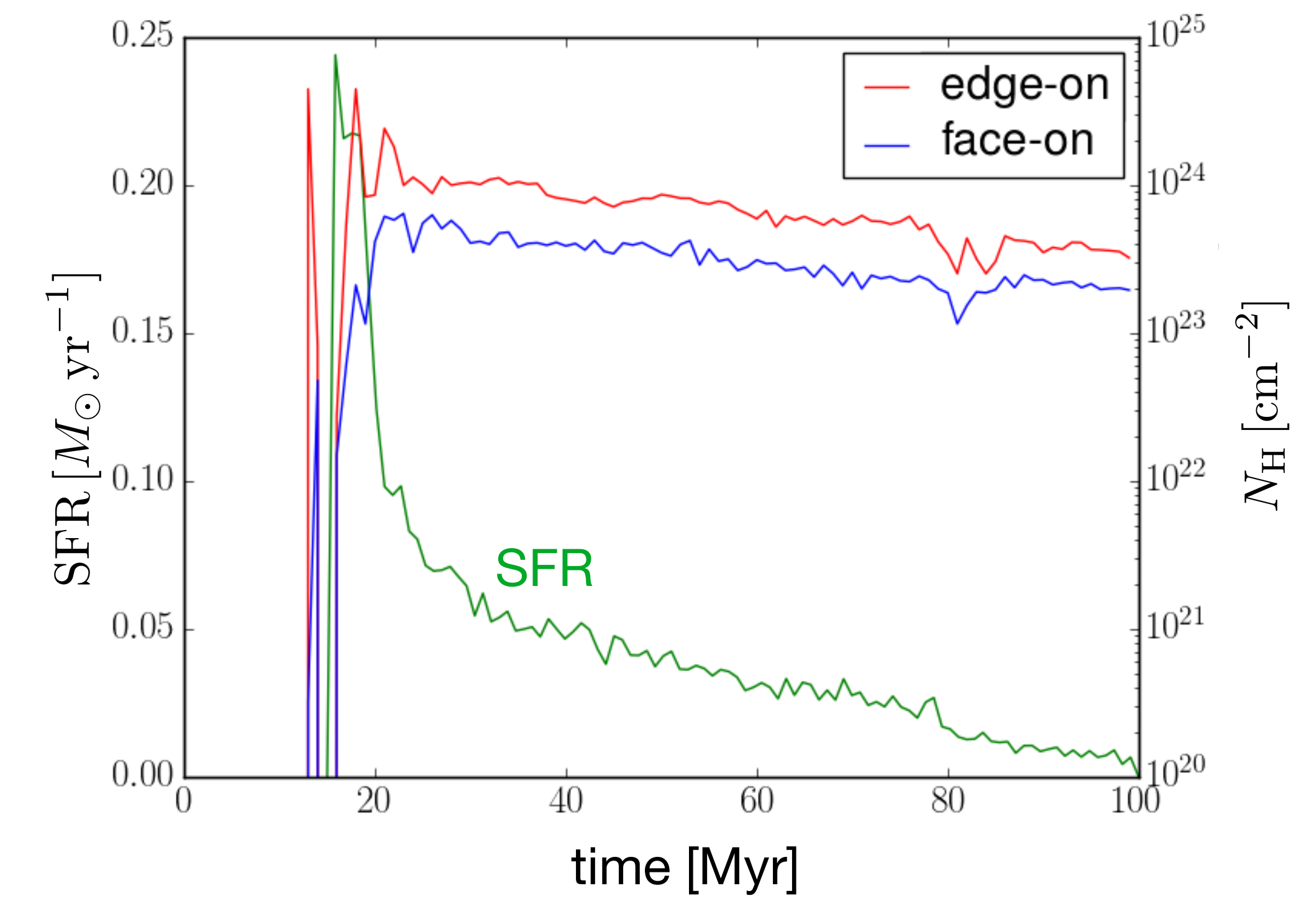}
\caption{Evolution of star formation rate (green line for left vertical axis) and column densities from two directions (red and blue lines) in Model AGN.}

\label{fig: 19}
\end{figure}

\begin{figure}[h]
\centering
\includegraphics[width = 12cm]{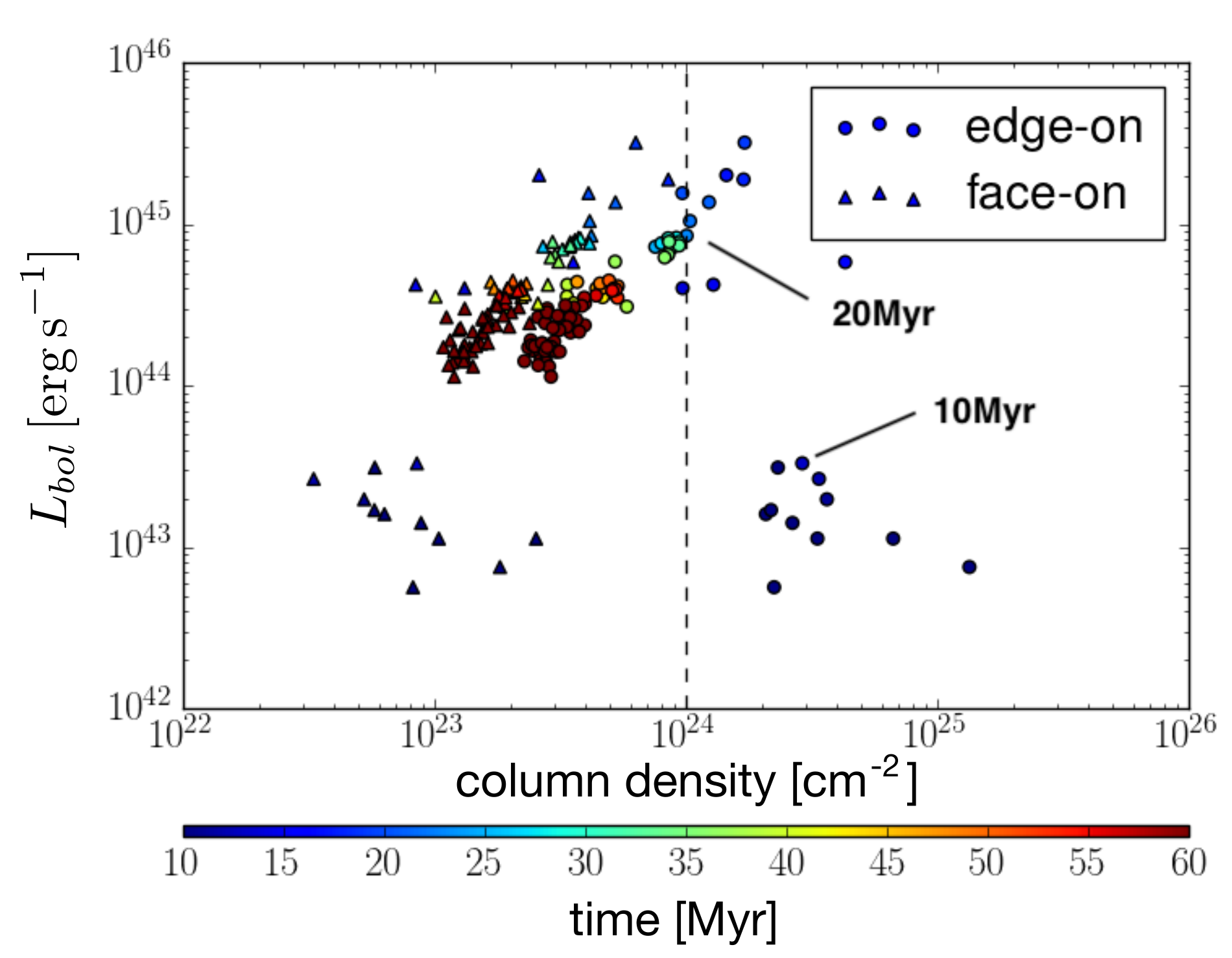}
\caption{Evolutionary track in model AGN on the plane of bolometric luminosity, {based on the mass accretion to the PBH,} and column densities 
{toward the PBH}.}

\label{fig: 20}
\end{figure}

\begin{figure}[h]
\centering
\includegraphics[width = 10cm]{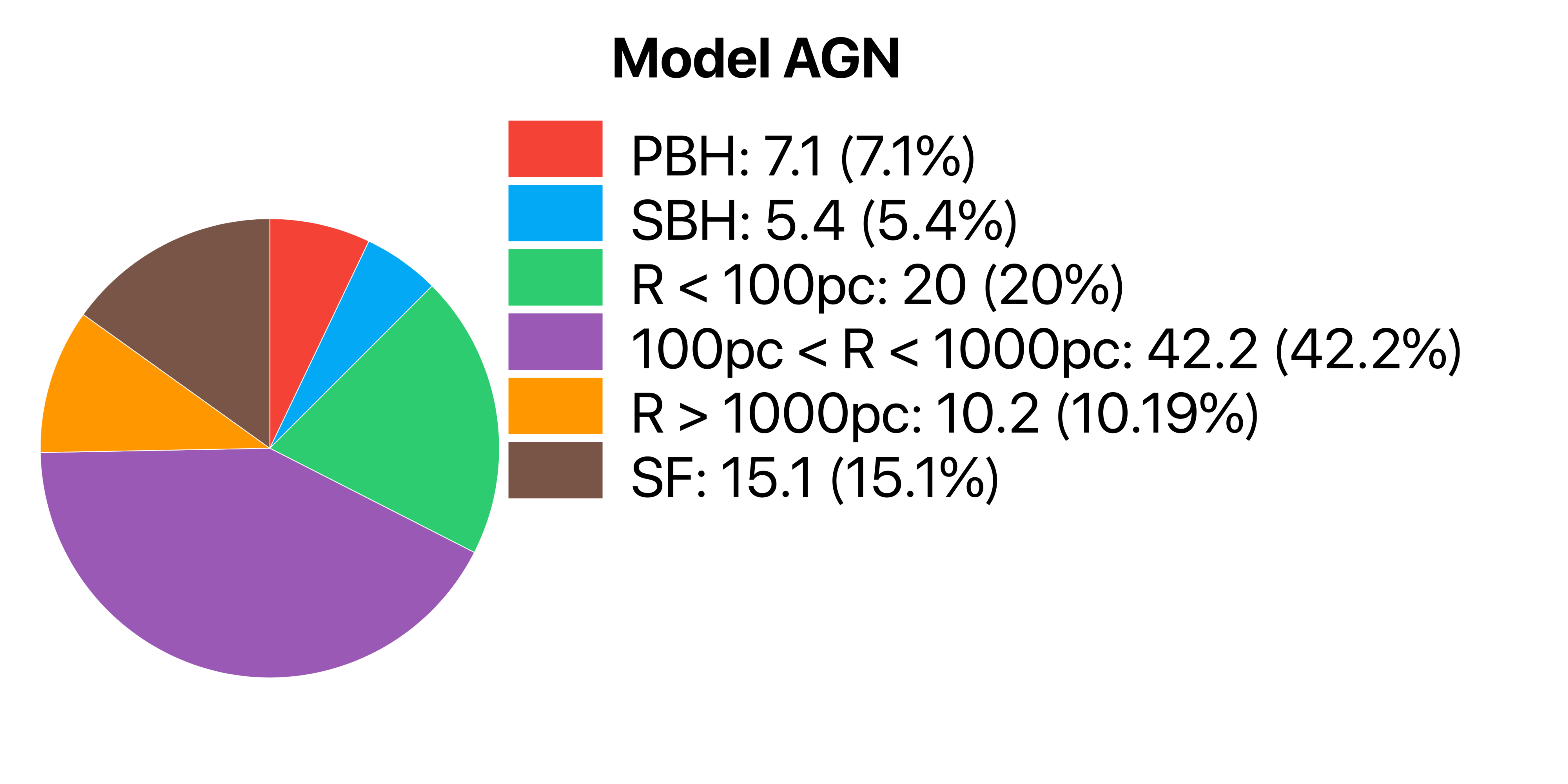} \\
\includegraphics[width = 10cm]{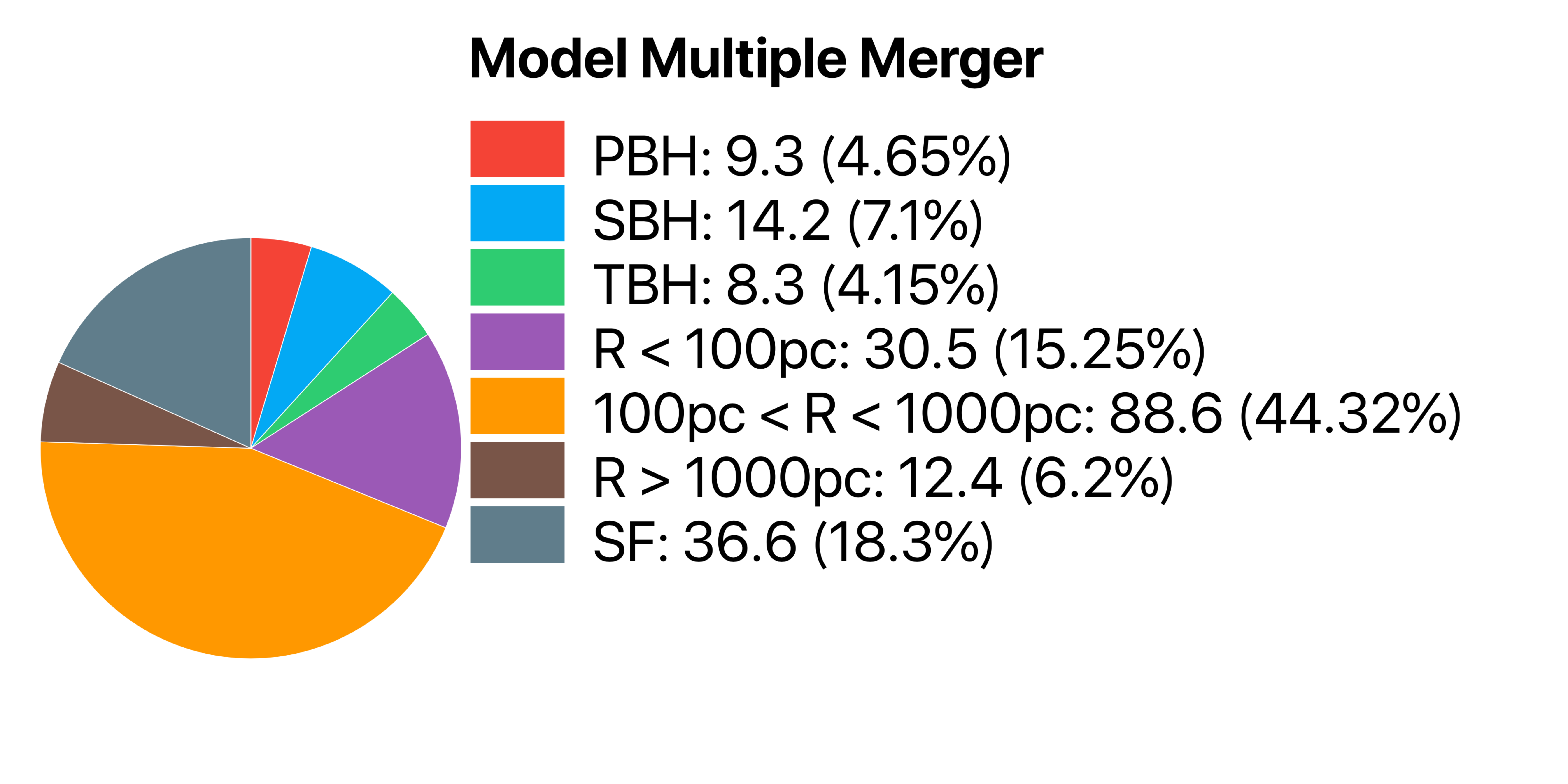}
\caption{Final destinations of merged gas in Models AGN and Multiples. Numbers are mass in $10^6 M_\odot$.}
\label{fig: 21}
\end{figure}

{
Infrared bright dust-obscured galaxies (DOGs) are a class of optically faint (the flux ratio between 24$\micron$ and $R$-band
is larger than 1000) massive galaxies in ultra-luminous infrared galaxies (ULIRGs) \citep{dey2008, toba2017}.
The origin of DOGs remains unclear, but \citet{narayanan2010} suggested that luminous DOGs are well-modeled by 
gas-rich major mergers. They also found that 24 $\micron$ bright sources are dominated by 
AGN emission, which is consistent with the X-ray properties of DOGs \citep[e.g.,][]{fiore2008, riguccini2019}.
Our result (Fig. \ref{fig: 20}) implies that the most luminous phase of the AGN ($t\sim 20$ Myr) 
is obscured by gas with a large column density ($N_{\rm H} \sim 10^{24}$ cm$^{-2}$).
Therefore, it is expected that most of the emissions from the nucleus ($\sim 10^{45}$ erg s$^{-1}$) should be absorbed by
the surrounding dusty gas and reemitted in the infrared band. This final phase of the gas-rich merger could correspond
to DOGs.
}

Although galactic mergers would be a primary mechanism for BH growth in galactic centers,
{the fraction of the gas consumed through accretion to the BHs is not clear.}
In Fig. \ref{fig: 21}, we show the final destinations of the gas in Model AGN and Model Multiple. 
In Model AGN, approximately 12\% of the gas falls to PBHs or SBHs, 20\% of the gas forms a circumnuclear gas cloud
within 100 pc, and 42\% forms a circumnuclear gas inside 1 kpc. Approximately 15\% of the gas is consumed by star formation.
The fraction of the gas accreted to BHs does not significantly increase, even in the multiple-merger model, 
holding steady at approximately 17\%. Approximately 60\% of the gas remains inside 1 kpc.

{\citet{ricci2017} studied the relation between AGN obscuration and galaxy mergers by analyzing the X-ray emission of 30 ULIRGs,
finding that in all AGNs in the late-merger stage, the column density toward the nucleus is $N_H > 10^{23}$ cm$^{-2}$.  They also 
 suggested that the obscuring material is effectively funneled from the galactic scale to the inner tens of parsecs during the late 
 stage of galaxy mergers. The compact obscuring material around the nucleus has been recently suggested in ULIRGs, known as 
 compact obscuring nuclei (CONS) \citep{sakamoto2010, costagliola2013, aalto2015, scoville2015}, {whose sizes are
 tens of parsecs, and their column density is $N_H > 10^{24}$ cm$^{-2}$}. These objects could correspond to the obscured AGNs appearing
  in the late-phase gas-rich mergers seen in our models.}


\section{Conclusions}
We studied the interactions between supermassive black holes (SMBHs) and the interstellar medium in the central sub-kpc region, using the N-body/SPH code ASURA \citep{saitoh2008, saitoh2009, saitoh2013}.
The numerical experiments aimed to understand the fate of the gas supplied by
mergers of two or more galaxies with SMBHs, and the efficiency of the BH growth resulting from
the gas supply to the galactic central region by mergers.

We found that the mass accretion rate to one SMBH
exceeds the Eddington ratio as the distance of two BHs rapidly
decreases.
However, this rapid accretion phase does not last more than 10 Myrs, and
it decreases to a
sub-Eddington value ($\sim$ 10\% of
the Eddington mass accretion rate).
{The rapid accretion is caused by the angular momentum transfer from the gas to
the stellar component, where gravitational torque dominates the torque created by the
turbulent pressure gradient.
The rapid accretion phase is followed by a quasi-steady accretion phase where
the angular momentum is redistributed not only by the gravitational torque,
 but also by the turbulent viscosity in the gas disk.}

We ran simulations with multiple-merger events and found that
the mass accretion rates to the BHs are similar to those in the first merger event.
The second rapid accretion phase was as short as the first one.
This implies that multiple mergers 
do not accelerate the mass accretion to BHs.
These features of the mass accretion resulting from mergers do not significantly depend on
the choice of the initial orbits of the merging system.

{
The AGN feedback could suppress the mass accretion to the BH. If this ``negative'' feedback works,
the AGN activity itself is suppressed, and the central BHs cannot grow through mass accretion.
Therefore, in reality, there should be an appropriate ``effective strength'' of the AGN feedback in 
the central region (e.g., $r <$  a few parsecs), for which
the mass accretion takes place even under the effect of the feedback.
We found in our numerical experiments that 
the AGN feedback and the mass accretion to BHs can coexist during galaxy mergers, 
 if the amount of the feedback energy to the central few pc is given as $(2 \times 10^{-4} - 2 \times 10^{-3} )\dot{M} c^2$, where 
 $\dot{M}$ is the accretion rate to $r= 1$ pc. 
The accretion rate, and therefore BH growth, is suppressed by $\sim$ 1/50 in the quasi-steady accretion phase
for $2 \times 10^{-2} \dot{M} c^2$.
}

{
The luminous phase of the AGN ($L_{bol} > 10^{45}$ erg s$^{-1}$) during the merger events is heavily obscured ($N_{H} > 10^{24}$ cm$^{-2}$) 
by the supplied gas, and the moderate AGN feedback does not alter this property.
The fraction of the gas that accretes to each BH is approximately 5--7\% of the supplied total gas mass ($10^8 M_\odot$),
 and 15--20\% of the gas forms a circumnuclear gas within 100 pc of the BH.}
We found that the BHs grow by a factor of 2 with respect to mass during a few merger events in $\sim 10^8$ yr if $10^8 M_\odot$ gas is supplied to the central kpc region.
The BH mass acquisition occurs mostly in the quasi-steady circumnuclear disk after the first rapid accretion, resulting from the completion of angular momentum transfer 
between the gas and stars. 
Star formation consumes approximately 15\% of the gas supplied by mergers,
and the rest forms a circumnuclear gas.
The efficiency of BH growth by mergers is approximately 10\% (i.e., only 1/10 of the supplied gas
accumulates to the central 1 pc)
for each event.
{This idealized situation 
implies that frequent mergers are necessary
for the continuous growth of BHs. However, in reality, the gas could also be
supplied to the central region, for example, through instabilities of the gas disk or by interaction with the galactic bar.
}
We found that the gas inside $r < 100$ pc mostly contributes to the large column density.
These properties may explain the recent findings of a high fraction of Compton-thick AGNs in merging systems and AGNs in dust-obscured galaxies (DOGs) \citep{dey2008, fiore2008, toba2017, riguccini2019}.

\acknowledgments
The authors would like to thank the anonymous referee for his/her many constructive
comments and helpful suggestions.
We are grateful to T. Saitoh for providing the ASURA code and his continuous support.
This work was supported by JSPS KAKENHI Grant Numbers 16H03959.


\listofchanges

\section*{Appendix:  Convergence tests}
{
In Fig. 16, we compare the mass accretion rates relative to the Eddington rate in two models with different numerical resolutions. 
PBH-20 and SBH-20 are the mass accretion rates in a model with $N_{SPH} = 2\times 10^5$,
and PBH-100 and SBH-100 are those in a model with $N_{SPH} = 10^6$ (i.e., 5 times greater
mass resolution).  Note that the model with $N_{SPH} = 2\times 10^5$ is different from 
Model hP in the sense that star formation is not included. Therefore, the gas disk in 
the SBH system initially collapses and as a result causes a large accretion rate at $t < 10$ Myr. 
The two results with different mass resolutions show good agreement.
After $t \sim 40 $ Myr, the accretion rate is a factor of two higher in the higher-resolution model.
The accretion rate for $ t > 65$ Myr also does not match. This would be caused by interaction between
the close binary of BHs and SPH particles. 
}

\begin{figure}[h]
\centering
\includegraphics[width = 10cm]{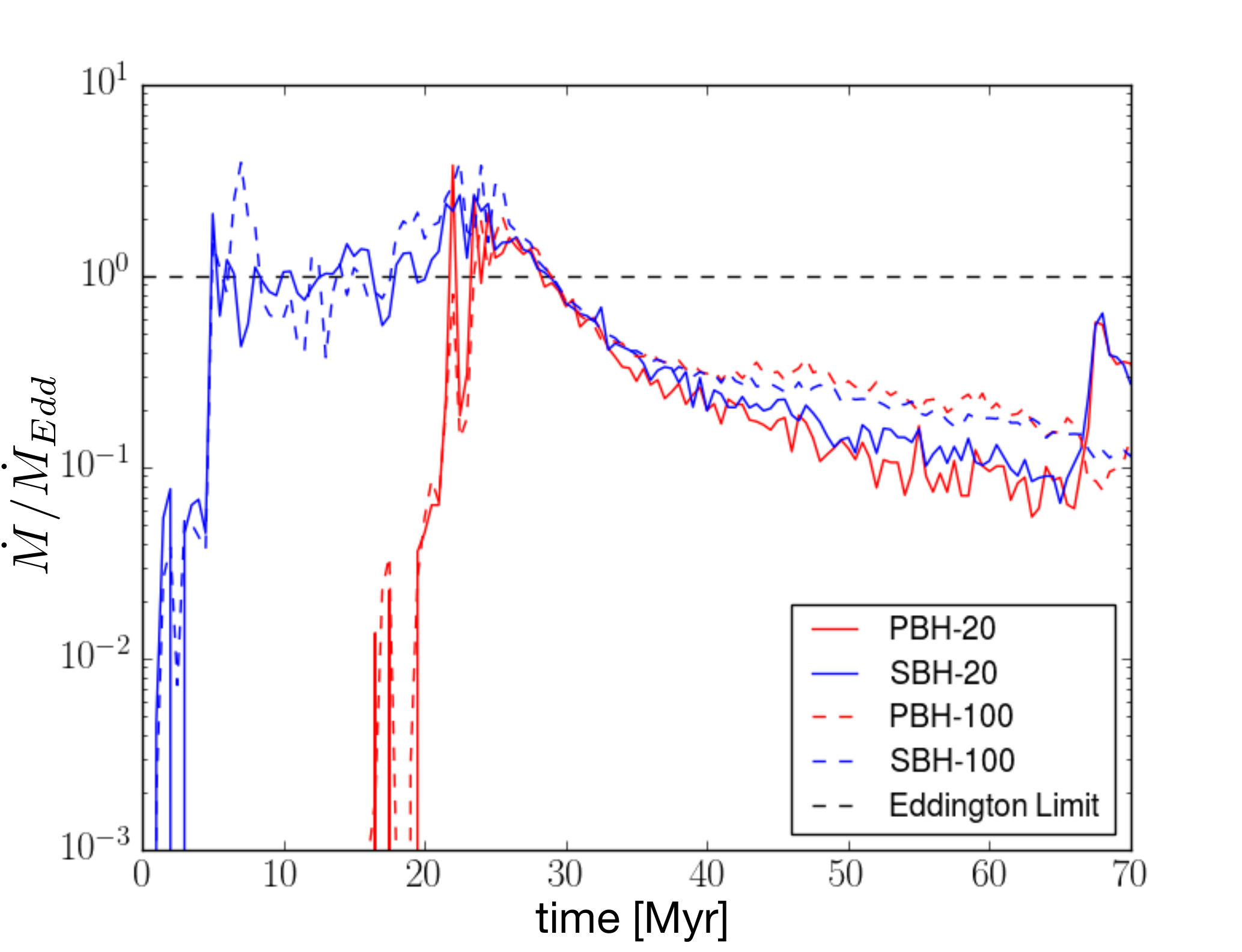} \\
\caption{{The accretion rates to PBH and SBH in two runs with different resolutions: $N_{SPH} = 2\times 10^5$ (solid lines) and $N_{SPH} = 10^6$ (dashed lines). 
 Here, star formation is not included in the gas disk.}}
\label{fig: converge}
\end{figure}


\begin{thebibliography}{}
\bibitem[Angl{\'e}s-Alc{\'a}zar et al.(2017)]{angles2017} Angl{\'e}s-Alc{\'a}zar, D., Faucher-Gigu{\`e}re, C.-A., Quataert, E., et al.\ 2017, \mnras, 472, L109 
\bibitem[Aalto et al.(2015)]{aalto2015} Aalto, S. 2015
\bibitem[Armitage \& Natarajan(2002)]{armitage2002} Armitage
\bibitem[Barnes \& Hernquist(1991)]{Barnes1991} Barnes, J.~E., \& Hernquist, L.~E.\ 1991, \apjl, 370, L65
\bibitem[Bigiel et al.(2008)]{bigiel2008} Bigiel et al.  
\bibitem[Blecha et al.(2018)]{2018MNRAS.478.3056B} Blecha, L., Snyder, G.~F., Satyapal, S., \& Ellison, S.~L.\ 2018, \mnras, 478, 3056
\bibitem[Buchner et al.(2015)]{buchner2015} Buchner, J., Georgakakis, A., Nandra, K., et al.\ 2015, \apj, 802, 89 
\bibitem[Cisternas et al.(2011)]{cisternas2011} Cisternas, M., Jahnke, K., Inskip, K.~J., et al.\ 2011, \apj, 726, 57
\bibitem[Costagliola et al.(2013)]{costagliola2013}  Costagliola 2013
\bibitem[Debuhr et al.(2011)]{Debuhr2011} Debuhr, J., Quataert, E., \& Ma, C.-P.\ 2011, \mnras, 412, 1341
\bibitem[del Valle \& Escala(2012)]{delvalle2012} del Valle ...2012
\bibitem[Dey et al.(2008)]{dey2008} Dey, A., Soifer, B.~T., Desai, V., et al.\ 2008, \apj, 677, 943 
\bibitem[Di Matteo et al.(2005)]{dimatteo2005} Di Matteo, T., Springel, V., \& Hernquist, L.\ 2005, \nat, 433, 604
\bibitem[Dotti et al.(2007)]{dotti2007} Dotti, M., Colpi, M., Haardt, F., et al.\ 2007, \mnras, 379, 956
\bibitem[Fan et al.(2001)]{Fan2001} Fan, X., Narayanan, V.~K., Lupton, R.~H., et al.\ 2001, \aj, 122, 2833
\bibitem[Fiore et al.(2008)]{fiore2008} Fiore, F., Grazian, A., Santini, P., et al.\ 2008, \apj, 672, 94
\bibitem[Gabor et al.(2009)]{gabor2009} Gabor, J.~M., Impey, C.~D., Jahnke, K., et al.\ 2009, \apj, 691, 705 
\bibitem[Goulding et al.(2018)]{goulding2018a} Goulding, A.~D., Zakamska, N.~L., Alexandroff, R.~M., et al.\ 2018, \apj, 856, 4 
\bibitem[Governato et al.(1994)]{governato1994} Governato, F., Colpi, M., \& Maraschi, L.\ 1994, \mnras, 271,  
\bibitem[Goulding et al.(2018)]{goulding2018b} Goulding, A.~D., Greene, J.~E., Bezanson, R., et al.\ 2018, \pasj, 70, S37 
\bibitem[Kazantzidis et al.(2005)]{kazantzidis2005} Kazantzidis 2005
\bibitem[Koss et al.(2016)]{Koss2016} Koss, M.~J., Assef, R., Balokovi{\'c}, M., et al.\ 2016, \apj, 825, 85
\bibitem[Hopkins(2009)]{2009AAS...21411101H} Hopkins, P.~F.\ 2009, American Astronomical Society Meeting Abstracts \#214, 214, 111.01
\bibitem[Hopkins et al.(2006)]{Hopkins2006} Hopkins, P.~F., Somerville, R.~S., Hernquist, L., et al.\ 2006, \apj, 652, 864
\bibitem[Hopkins et al.(2011)]{Hopkins2011} Hopkins, P.~F., Quataert, E., \& Murray, N.\ 2011, \mnras, 417, 950
\bibitem[Hopkins \& Quataert(2011)]{hopkins2011} Hopkins, P.~F., \& Quataert, E.\ 2011, \mnras, 415, 1027
\bibitem[Kocevski et al.(2012)]{kocevski2012} Kocevski, D.~D., Faber, S.~M., Mozena, M., et al.\ 2012, \apj, 744, 148
\bibitem[Lanzuisi et al.(2015)]{Lanzuisi2015} Lanzuisi, G., Perna, M., Delvecchio, I., et al.\ 2015, \aap, 578, A120
\bibitem[Lin \& Pringle(1987)]{lin_pringle1987} Lin, Pringle 1987
\bibitem[Mayer et al.(2007)]{mayer2007} Mayer, L., Kazantzidis, S., Madau, P., et al.\ 2007, Science, 316, 1874
\bibitem[Marconi, \& Hunt(2003)]{marconi2003} Marconi, A., \& Hunt, L.~K.\ 2003, \apjl, 589, L21
\bibitem[Mihos \& Hernquist(1996)]{mihos1996} Mihos, Hernquist 1996
\bibitem[Mortlock et al.(2011)]{Mortlock2011} Mortlock, D., Warren, S., Patel, M., et al.\ 2011, Galaxy Formation, 88
\bibitem[Narayanan et al.(2010)]{narayanan2010} Narayanan, D., Dey, A., Hayward, C.~C., et al.\ 2010, \mnras, 407, 1701
\bibitem[Negri et al.(2018)]{Negri2018} Negri, M., Turchi, M., Chatterjee, R., \& Bertoldi, N.\ 2018, arXiv:1803.07274
\bibitem[Okamoto et al.(2008)]{okamoto2008} Okamoto, T., Nemmen, R.~S., \& Bower, R.~G.\ 2008, \mnras, 385, 161 
\bibitem[Ricci et al.(2017)]{ricci2017} Ricci, C., Bauer, F.~E., Treister, E., et al.\ 2017, \mnras, 468, 1273
\bibitem[Riguccini et al.(2019)]{riguccini2019} Riguccini, L.~A., Treister, E., Men{\'e}ndez-Delmestre, K., et al.\ 2019, \aj, 157, 233
\bibitem[Sakamoto et al.(2010)]{sakamoto2010}  Sakamoto
\bibitem[Saitoh et al.(2008)]{saitoh2008} Saitoh, T.~R., Daisaka, H., Kokubo, E., et al.\ 2008, \pasj, 60, 667
\bibitem[Saitoh et al.(2009)]{saitoh2009} Saitoh, T.~R., Daisaka, H., Kokubo, E., et al.\ 2009, \pasj, 61, 481
\bibitem[Saitoh \& Wada(2004)]{saitoh2004} Saitoh, T.~R., \& Wada, K.\ 2004, \apjl, 615, L93 
\bibitem[Saitoh, \& Makino(2013)]{saitoh2013} Saitoh, T.~R., \& Makino, J.\ 2013, \apj, 768, 44
\bibitem[Schawinski et al.(2012)]{2012MNRAS.425L..61S} Schawinski, K., Simmons, B.~D., Urry, C.~M., Treister, E., \& Glikman, E.\ 2012, \mnras, 425, L61
\bibitem[Scoville et al.(2015)]{scoville2015} Scoville 2015
\bibitem[Schmidt(1959)]{schmidt1959} Schmidt, M.\ 1959, \apj, 129, 243
\bibitem[Souza Lima et al.(2017)]{Souza2017} Souza Lima, R., Mayer, L., Capelo, P.~R., \& Bellovary, J.~M.\ 2017, \apj, 838, 13
\bibitem[Springel \& Hernquist(2003)]{Springel2003} Springel, V., \& Hernquist, L.\ 2003, Astrophysical Supercomputing using Particle Simulations, 208, 273
\bibitem[Springel et al.(2005)]{Springel2005} Springel, V., Di Matteo, T., \& Hernquist, L.\ 2005, \mnras, 361, 776
\bibitem[Taniguchi \& Wada(1996)]{taniguchi1996} Taniguchi, Y., \& Wada, K.\ 1996, \apj, 469, 581
\bibitem[Trebitsch et al.(2019)]{trebitsch2019} Trebitsch, M., Volonteri, M., \& Dubois, Y.\ 2019, \mnras, 487, 819
\bibitem[Toba et al.(2017)]{toba2017} Toba, Y., Nagao, T., Kajisawa, M., et al.\ 2017, \apj, 835, 36 
\bibitem[Treister et al.(2012)]{treister2012} Treister, E., Schawinski, K., Urry, C.~M., \& Simmons, B.~D.\ 2012, \apjl, 758, L39
\bibitem[Tremmel et al.(2018)]{tremmel2018} Tremmel, M., .. 2018
\bibitem[Prieto et al.(2017)]{Prieto2017} Prieto, J., Escala, A., Volonteri, M., \& Dubois, Y.\ 2017, \apj, 836, 216
\bibitem[Ramos Almeida \& Ricci(2017)]{ramos-almeida2017} Ramos Almeida, C., \& Ricci, C.\ 2017, Nature Astronomy, 1, 679 
\bibitem[Villforth et al.(2014)]{2014MNRAS.439.3342V} Villforth, C., Hamann, F., Rosario, D.~J., et al.\ 2014, \mnras, 439, 3342
\bibitem[Villforth et al.(2017)]{2017MNRAS.466..812V} Villforth, C., Hamilton, T., Pawlik, M.~M., et al.\ 2017, \mnras, 466, 812
\bibitem[Wada et al.(2009)]{wada2009} Wada, K., Papadopoulos, P.~P., \& Spaans, M.\ 2009, \apj, 702, 63 
\bibitem[Wada(2015)]{wada2015} Wada, K.  2015
\bibitem[Weston et al.(2017)]{weston2017} Weston, M.~E., McIntosh, D.~H., Brodwin, M., et al.\ 2017, \mnras, 464, 3882 
\end{thebibliography}
\end{document}